\documentclass{article}
\usepackage{graphicx}
\usepackage{here}

\def\beq{\begin{equation}}
\def\eeq{\end{equation}}
\def\bea{\begin{eqnarray}}
\def\eea{\end{eqnarray}}
\def\bq{\begin{quote}}
\def\eq{\end{quote}}

\def\nnb{\nonumber}
\def\ga{\left(}
\def\dr{\right)}
\def\aga{\left\{}
\def\adr{\right\}}
\def\lb{\lbrack}
\def\rb{\rbrack}
\def\rar{\rightarrow}
\def\lrar{\Longrightarrow}
\def\nnb{\nonumber}
\def\la{\langle}
\def\ra{\rangle}
\def\nin{\noindent}
\def\ba{\begin{array}}
\def\ea{\end{array}}

\def\b{\bullet}

\def\als{\alpha_s}
\def\as{\ga\frac{\bar{\alpha_s}}{\pi}\dr}
\def\asr{\ga\frac{{\alpha_s}}{\pi}\dr}
\def\gg{ \la\alpha_s G^2 \ra}
\def\ggg{g^3f_{abc}\la G^aG^bG^c \ra}

\def\lnu{\log{-\frac{q^2}{\nu^2}}}
\begin{document}

\pagestyle{empty}
\begin{flushright}
PM 96/37\\
hep-ph/9612448
\end{flushright}
\vspace*{1cm}
\begin{center}
\section*{Masses, Decays and Mixings of Gluonia in QCD }

\vspace*{1.0cm}
{\bf S. Narison}\footnote{email address: narison@lpm.univ-montp2.fr} \\
\vspace{0.3cm}
Laboratoire de Physique Math\'ematique et Th\'{e}orique\\
Universit\'e de Montpellier II\\
Place Eug\`ene Bataillon\\
34095 - Montpellier Cedex 05, France\\
\vspace*{1.0cm}
{\bf Abstract} \\ \end{center}
\vspace*{2mm}

\noindent
We compute the masses and decay widths of the gluonia using QCD
spectral sum rules and low-energy theorems. In the scalar
sector, one finds a gluonium having a mass
 $M_G=(1.5\pm 0.2)$ GeV, which decays mainly into the $U(1)_A$ channels
$\eta\eta'$ and $4\pi^0$. However,
for a consistency of the whole approach, one needs
broad-low mass gluonia (the $\sigma_B$ and its radial excitation), 
which couple strongly to the quark degrees of freedom similarly to 
the $\eta'$ of the $U(1)_A$ sector. Combining
these results with the ones for the
$\bar qq$ quarkonia, we present maximal
gluonium-quarkonium mixing schemes, which
can provide quite a good description of the 
complex spectra and various decay widths of the observed scalar mesons
$\sigma(1.),~ f_0(0.98),~f_0(1.37),~f_0(1.5)$ and $f_J(1.71)$. 
In the tensor sector, the gluonium mass is found to be
 $M_T\simeq (2.0\pm 0.1)$ GeV, which
makes the $\zeta(2.2)$ a good $2^{++}$ gluonium candidate, even though
we expect a rich population of $2^{++}$ gluonia in this region.
In the pseudoscalar
channel, the gluonium mass is
found to be $M_P\simeq (2.05\pm 0.19)$ GeV, while we also show that
the $E/\iota$(1.44) couples more weakly to the gluonic current
than the $\eta'(0.96)$, which can 
favour its interpretation as
the first radial excitation of the $\eta'$(0.96).
\vspace*{3cm} 
\begin{flushleft}
PM 96/37\\
hep-ph/9612448\\
December 1996
\end{flushleft}
\vfill\eject
\setcounter{page}{1}
\pagestyle{plain}
\section{Introduction}
There is now clear evidence from many processes
that QCD is the theory of strong interactions. All hard 
processes satisfy the asymptotic freedom property of QCD, while the
complicated hadron properties can be explained by different 
non-perturbative methods (effective Lagrangian, lattice calculations, 
QCD spectral sum rule (QSSR),...). However, in addition to the well-known
mesons and baryons bound states, one of the main consequences of the 
QCD theory is the possible existence of the gluon bound states 
(gluonia or glueballs) or/and of a gluon continuum.
The theoretical interest 
in the gluonia sector has started a long time ago, 
since the pionneering work of Fritzsch and Gell-Mann \cite{FRITZ}, 
 as shown by
the long list of publications on this topic. However,
despite these different efforts, the theoretical and experimental status 
remains unclear. From the theoretical point of view, this is due to our poor
control of the gluon dynamics. In particular, there is not yet any convincing
dynamical approach that computes the mixing of the gluonia with quarkonium
states (see however \cite{PAK,BRAMON}),
 which is necessary if one wishes to make contact with the experimental
data. At this level, only phenomenological scheme is available, where the
mixing angle is only fitted from the data. From the experimental point of view,
the difficulty also arises in the same way, as the observed resonances can be
a mixing between gluonia and quarkonia. However, some selected processes 
such as
the $J/\psi$ radiative decays can favour more the production of gluonia than of
quarkonia, while the measurement of the two-photon widths has been used for
a long time as a good gluonia signature. However, as proposed some years ago 
in \cite{CHAN}, a good signature 
for the presence of the gluonia can be obtained from the ratio of the previous
two quantities referred to as ``stickiness". Another 
possibility for signing the nature of an almost pure gluonium is the 
measurements of 
its width into the U(1)-like channels: $\eta'\eta'$, $\eta\eta'$, and $4\pi^0$, which 
are expected to be more dominant than, for instance, the one into pair of pions, 
leading \cite{VENEZIA,GERS} to the conclusions that the G(1.6) observed by 
the GAMS group \cite{GAMS} is an almost pure glueball state. The different 
experimental progress done during these 
last few years \cite{LANDUA}, though not yet very conclusive , 
is encouraging for improving the theoretical predictions. Some recent efforts 
have been accomplished in this direction from lattice calculations 
\cite{SHARPE}--\cite{WEIN}. However, the ``apparent" disagreements of
these lattice results can simply reflect the true systematic
error of the estimates from
this approach. Recently, some QCD inequalities among the gluonia
masses have been derived \cite{WEST}.
Some qualitative phenomenological attempts \cite{CLOSE} have also been 
proposed
for explaining the nature of the
scalar meson seen by the Crystal Barrel collaboration \cite{BARREL}. 
Motivated by these different steps forward, we plan to give in this paper
an almost complete scheme and an update of the predictions 
from QSSR \`{a} la
SVZ \cite{SVZ} \footnote{For a recent review on the sum rules, 
see e.g. \cite{SNB}}, where, in my
opinion, the results have often been misinterpreted in the literature, and
in some cases ignored. 
A long list of sum rules results exists in the literature 
\cite{NSVZ}--\cite{BOOS}, which needs to be updated
because of the progress accomplished in QCD
during the last few years. This is the aim of this work.
The paper is organized as follows: \\
In section 2, we give a general discussion on the gluonia and the 
classification of different currents.
In section 3, we give a short introduction to the method
of QSSR.
In sections 4 and 5, we update mainly the work of \cite{VENEZIA} (hereafter
referred to as NV) for the $0^{++}$ sector by improving the QSSR
predictions on the masses and couplings of the gluonia, 
and by extending the uses of some low-energy theorems (LET)
for their decay widths, into new predictions for their decay into $4\pi^0$.
In section 6, we compute the properties (masses and widths) of the scalar quarkonium.
In section 7, as in \cite{BRAMON2} (hereafter
referred to as BN), we use some maximal quarkonium-gluonium mixing schemes 
for explaining the
scalar mesons data below and above 1 GeV.
In section 8, we discuss the QSSR estimate of the 2$^{++}$ gluonium
mass and coupling, which is an update of the work of \cite{SNG}
(hereafter referred to as SN). 
We close this section by giving the widths of the $2^{++}$ gluonium
using a meson-gluonium mixing scheme and LET as in \cite{BRAMON}.\\
In section 9, we discuss the QSSR estimate of the $0^{-+}$ gluonium
mass and coupling (update of \cite{SNG}). We use the quarkonium-gluonium
mixing scheme for
predicting the $\gamma\gamma$ and $\rho\gamma$ decay of the
pseudoscalar gluonium. An attempt to explain the property of the 
$E/\iota$ (1.44) is
given. A summary of our results is given in the two tables in section 10.
\section {The gluonic currents}
In this paper, we shall consider the lowest-dimension gluonic currents
that can be built from the gluon fields and are gauge-invariant:\\
\bea
J_s&=& \beta(\alpha_s) G_{\alpha\beta}G^{\alpha\beta},\nnb\\
\theta^g_{\mu\nu}&=&-G_{\mu}^{\alpha}G_{\nu\alpha}+\frac{1}{4}g_{\mu\nu}
G_{\alpha\beta}G^{\alpha\beta}, \nnb\\
Q(x)&=&\ga \frac{\alpha_s}{8\pi}\dr \mbox{tr}~G_{\alpha\beta}
\tilde{G}^{\alpha\beta},
\eea
where the sum over colour is understood and in our notations \cite{SNB}
\footnote{The corresponding differential equation for
the running coupling is $d\bar{\alpha}_s/dt=
\bar{\alpha}_s \beta(\bar{\alpha}_s)$ where $ t\equiv 1/2\log(-q^2/\nu^2)$.}:
\bea
\beta(\alpha_s)&=&\beta_1\as+\beta_2\as^2+{\cal{O}}(\alpha_s^3):
~~~~\beta_1=-\frac{11}{2}+\frac{n_f}{3},
~~~~~~~\beta_2=-\frac{1}{4}\ga 51-
\frac{19}{3}n_f\dr ,\nnb\\
\tilde{G}_{\mu\nu}&=&\frac{1}{2}\epsilon_{\mu\nu\alpha\beta}{G}^{\alpha\beta}.
\eea
 These currents have respectively the 
quantum numbers of the $J^{PC}= 0^{++},~2^{++}$ and $0^{-+}$ 
gluonia \footnote{We shall not consider the pseudotensor $2^{-+}$ 
in this paper. A QSSR analysis of this channel is in progress.}, 
which
are familiar in QCD. The former two enter into the QCD energy-momentum 
tensor
$\theta_{\mu\nu}$, while the later is the U(1)$_A$ axial-anomaly current. 
The
renormalizations and corresponding anomalous
dimension of the scalar and pseudoscalar currents have been studied
in QCD \cite{TARRACH}, where renormalization group invariant quantities 
have been built. In the approximations, at which we are working, we can ignore
renormalization effects relevant at higher orders of the $\als$ radiative
corrections. Besides the sum rules analysis of the corresponding two-point 
correlators, we shall also study the gluonia properties using 
some LET, while more phenomenological mixing schemes will be presented
to explain the data.
\section{QCD spectral sum rules}

\nin
The analysis of the gluonia masses and couplings will be
done using the method of
QSSR. In so doing, we shall work with the generic 
two-point correlator: 
\beq
\psi_G(q^2) \equiv i \int d^4x ~e^{iqx} \
\la 0\vert {\cal T}
J_G(x)
\ga J_G(0)\dr ^\dagger \vert 0 \ra ,
\eeq
built from the previous gluonic currents $J_G(x)$.
Thanks to its analyticity property, the correlator obeys the well-known
K\"allen--Lehmann dispersion relation:
\beq
\psi_G (q^2) = 
\int_{0}^{\infty} \frac{dt}{t-q^2-i\epsilon}
~\frac{1}{\pi}~\mbox{Im}  \psi_G(t) ~ + ...,
\eeq
where ... represent subtraction points, which are
polynomials in the $q^2$-variable. This $sum~rule$
expresses in a clear way the {\it duality} between the integral involving the 
spectral function Im$ \psi_G(t)$ (which can be measured experimentally), 
and the full correlator $\psi_G(q^2)$, 
which can be calculated directly in
QCD using perturbation theory and the Wilson expansion, provided 
that $-q^2$ is much greater than $\Lambda^2$.\\
\subsection{The Operator Product Expansion}

\nin
By adding to the usual perturbative expression of the correlator,
the non-perturbative contributions, as parametrized by the vacuum
condensates of higher and higher dimensions in the OPE \cite{SVZ},
the two-point correlator reads in QCD:
\beq
\psi_G(q^2)
\simeq \sum_{D=0,2,4,...}\frac{1}{\ga -q^2 \dr^{D/2}} 
.\sum_{dim O=D} C^{(J)}(q^2,\nu)\la O(\nu)\ra,
\eeq
where $\nu$ is an arbitrary scale that separates the long- and
short-distance dynamics; $C^{(J)}$ are the Wilson coefficients calculable
in perturbative QCD by means of Feynman diagrams techniques.
In the present analysis, we shall limit ourselves
to the computation of the gluonia masses
in the massless quark limit 
\footnote{Quark mass corrections
which will appear at higher order through internal fermion loops in the OPE
at the order $\alpha_s^3 (\bar{m}^2_s/q^2)\log(-q^2/\nu^2)$, are expected to be 
tiny at the
scale (see next sections) where the sum rules are optimized ($\bar{m}_s$ being the running
quark mass).}, which we shall compare with the results 
obtained from other methods such as lattice calculations and QCD
inequalities. The OPE is shown diagramatically in Fig. 1:\\
\begin{figure}[H]
\begin{center}
\includegraphics[width=6cm]{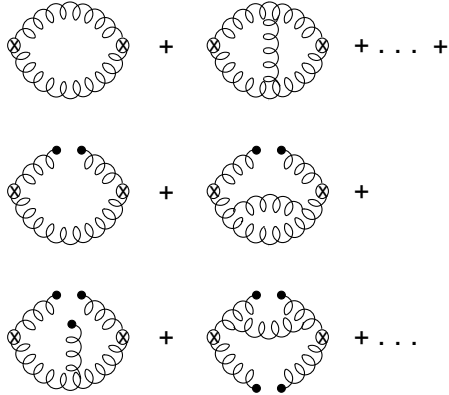}
\caption{ Diagrammatic representation of the OPE of the gluonic
two-point correlator}
\end{center}
\end{figure}

\nin
$\b$ The $D=0$ operator corresponds to the usual case of the na\"\i ve 
perturbative 
contribution. In the massless light quark limit , some additionnal $D=2$ terms not 
included in the OPE may eventually
manifest after the resummation of the QCD perturbative series \cite{ALT}--\cite{VAIN}. 
However, due to the unclear present status of this contribution, and to the inaccurate 
quantitative control of the contribution of this term from ultraviolet 
renormalons calculus, we shall not explicitly
consider this effect in our analysis. Alternatively,
we estimate, like in the case of $\tau$ decay
 \cite{BNP}--\cite{PICH}, the next unkown higher-order 
correction using a geometrical growth of the QCD series, which value is in agreement
with the one from experimental measurements \cite{LEDIBERDER} and from an estimate
 \cite{KATA} based on a $RS$-invariant
approach \footnote{The $RS$-invariant approach can be applied in the present case of gluonia
as the corresponding two-point function starts to lowest order as $\alpha_s^2$, such that the
unknown correction to be estimated is at order $\alpha_s^4$.}
within the principle of minimal sensitivity \cite{STEVENSON}. Then, we argue, like in
the case of $\tau$ decay, that
the errors induced by the truncation of the QCD series are given by this last term
(theorem of divergent series \cite{HARDY}), which we have multiplied by a factor 2
as in \cite{PICH} for a conservative estimate. The errors estimated in this way
are comparable in strength with some of the previous UV renormalon 
estimates, and should already include in it this renormalon contribution
(manifestation of the resummed QCD series).
However, as mentioned previously, the UV renormalon estimate
should be considered to be very qualitative at the present status
of the art (dominance of the second chain of renormalons \cite{VAIN},...).   
\\
$\b$ The $D=4$ dimension operators, which will give the dominant contributions
in the chiral limit $m_i=0$ is the gluonic condensate $\la\alpha_s G^2 \ra$,
introduced by SVZ \cite{SVZ}, and which has been estimated recently from the
$e^+e^-\rar~I=1$ hadron data \cite{SNL} and from the heavy
quarks mass splittings \cite{SNH}:
\beq
 \la\alpha_s G^2 \ra \simeq (0.07\pm 0.01)~\mbox{GeV}^4,
\eeq
in agreement with different post-SVZ estimates quoted in these papers.\\
$\b$ The $D=6$ contribution is dominated by the triple gluon condensate
$gf_{abc}\la G^aG^bG^c \ra$, whose direct extraction from
the data is still lacking. However, one can estimate its approximate 
value from the dilute  gas instanton model \cite{SVZ}:
 \beq
 g^3f_{abc}\la G^aG^bG^c \ra\approx (1.5\pm 0.5) \mbox{GeV}^2)\la\alpha_s G^2 \ra~,
\eeq
where we have given the error in such a way that the estimate is valid within a factor 2.

\nin
$\b$ The validity of the SVZ expansion has been intensively studied in
the theoretical literature \cite{SNB}, while
the good agreement for the values of the QCD coupling 
$\alpha_s$ from $\tau$ decay \cite{BNP}--\cite{PICH} and LEP data
\footnote{New data from deep inelastic scattering give now a value of $\alpha_s$
in a better agreement with the two former \cite{BETHKE}.},
the observation of the running of $\alpha_s$ at low-scale \cite{NEUBERT},
and the agreement of the measured QCD condensates from semi-inclusive
tau decays and spectral moments \cite{EXP} with the ones
from phenomenological fits \cite{SNL,SNH}, can be considered as
its phenomenological confirmation.\\ 
\nin
$\b$ In the case of gluonia studied in this paper, it has also been argued 
\cite{NSVZ}, using the
dilute gas approximation, that in some channels, instanton plus 
anti-instanton effects manifest themselves
as higher dimension $(D=11)$ operators. On the other hand, its 
quantitative estimate is quite inaccurate because of
 the great sensitivity of the result on
the QCD scale $\Lambda$, and on some other less controllable parameters and
coefficients. However, at the scale (gluonia scale) where the following sum rules
are optimized, which is much higher
than the usual case of the $\rho$ meson, we consider that
one can safely omit these terms like any other 
higher-dimensional operators beyond $D=8.$

\subsection{The spectral function and its experimental measurement}
The experimental measurement of the spectral function is best illustrated
in the case of the flavour-diagonal 
light quark vector current, where the spectral function Im$\Pi$(t) can
be related to the $e^+e^-$ into $I=1$ hadrons data via the optical
theorem as:
\beq
\sigma (e^+e^-\rar I=1~\mbox{hadrons})=
\frac{4\pi^2\alpha}{t}e^2~\frac{1}{\pi}~
\mbox{Im}  \Pi(t). 
\eeq
One can also relate the spectral function to the leptonic width of 
the $\rho$ resonance:
\beq
\Gamma_{\rho\rar e^+e^-}\simeq \frac{2}{3}\pi \alpha^2\frac{M_\rho}
{2\gamma^2_\rho},
\eeq
via the meson coupling to the electromagnetic current:
\beq
\la 0|J^\mu|\rho\ra =\frac{M^2_\rho}{2\gamma_\rho}\epsilon^\mu,
\eeq
within a vector meson dominance assumption. More generally, one can introduce
the decay constant $f_G$ analogous to $f_\pi=93.3$ MeV:
\beq
\la 0|J_G|G\ra =\sqrt{2}f_GM_G^2~...,
\eeq
where ...~represent the Lorentz structure of the matrix elements.
\subsection{The form of the sum rules} 
\nin
On the phenomenological side, the improvement of the 
previous dispersion relation comes from the 
uses of
an infinite number of derivatives and infinite values of $q^2$, 
but keeping their ratio fixed as $\tau \equiv n/q^2$. In this
way, one obtains the Laplace or Borel or exponential sum rules \cite{SVZ,BB,SR}:
\beq
{\cal L}_G(\tau)
= \int_{t_\leq}^{\infty} {dt}~\mbox{exp}(-t\tau)
~\frac{1}{\pi}~\mbox{Im} \psi_G(t),
\eeq
where $t_\leq$ is the hadronic threshold. The advantage of this sum 
rule with respect to the previous dispersion relation is the
presence of the exponential weight factor, which enhances the 
contribution of the lowest resonance and low-energy region
accessible experimentally. For the QCD side, this procedure has
eliminated the ambiguity carried by subtraction constants,
arbitrary polynomial
in $q^2$, and has improved the convergence of
the OPE by the presence of the factorial dumping factor for each
condensates of given dimensions. 

\nin
The ratio of sum rules:
\beq
{\cal R}_G \equiv -\frac{d}{d\tau} \log {{\cal L}_G},
\eeq
 or its slight modification, is a useful quantity to work with,
 in the determination of the resonance mass, as it is equal to the 
mass squared, in  
 the simple duality ansatz parametrization:
\beq
``\mbox{one resonance}"\delta(t-M^2_R)
 \ + \
 ``\mbox{QCD continuum}" \Theta (t-t_c),
\eeq
of the spectral function, where the
resonance enters by its coupling to the quark current; 
$t_c$ is
the continuum threshold which is, like the 
sum rule variable $\tau$, an  a priori arbitrary 
parameter. As tested in the meson channels, this
parametrization gives a good description of the spectral integral, in 
the sum rule analysis. In some cases, we shall also use finite energy 
sum Rule (FESR) \cite{LAUNER, KRAS}:
\beq
{\cal M}_G^{(n)}
= \int_{t_\leq}^{t_c} {dt}~t^n
~\frac{1}{\pi}~\mbox{Im} \psi_G(t),~~~~~~~~~~
{\cal R}_G^{(n)} \equiv \frac{{\cal M}_G^{(n)}}{{\cal M}_G^{(n+1)}},
\eeq
where $n$ is an integer, in order to double check the estimate 
obtained from the Laplace sum rules.

\subsection{Conservative optimization criteria}
Different optimization criteria are proposed in the literature,
which, to my opinion, complete one another, if used carefully.
The {\it sum rule window} of SVZ is a compromise region
where, at the same time, the OPE makes sense while the spectral
integral is still dominated by the lowest 
resonance. This is indeed
satisfied when the Laplace sum rule presents a minimum in
$\tau$, where there is an equilibrium between the
non-perturbative and high-energy region effects.
However, this criterion is not yet sufficient as the value of this
minimum in $\tau$ can still be greatly affected by the value
of the continuum threshold $t_c$. 
The needed extra condition is to find the region where
the result has also a minimal sensitivity on the change of the
$t_c$ values ($t_c$ stability).
The $t_c$ values obtained in this way are about the same as the one from
the so-called heat evolution test of the local duality 
FESR \cite{LAUNER}. However, in some cases, this $t_c$
value is too high, compared with the mass of the
observed radial excitation, and the procedure
tends to overestimate the predictions. More precisely, the result obtained
in this way can be considered as a phenomenological
upper limit.
Therefore, in order to have a {\it conservative} prediction from 
the sum rules method, one can consider the value of $t_c$ at which one
starts to have a $\tau$-stability up to where one
has a $t_c$ stability. In case there is no $t_c$ stability
nor FESR constraint on $t_c$, one can consider that the
prediction is still unreliable. In this paper, we shall limit ourselves to
extracting the results satisfying the $\tau$ (Laplace) or $n$ (FESR)
 and $t_c$ stability criteria.\footnote{Many results in the literature
on QCD spectral 
sum rules literature are obtained using  only the first condition.}  
\section{Mass and decay constant of the $0^{++}$ scalar gluonia}
\subsection{The gluonium two-point correlator in QCD}
We shall be concerned with the correlator:
\beq
\psi_s(q^2)\equiv 16 i\int d^4x~ e^{iqx}~\la 0|{\cal T}\theta^\mu_\mu(x)
\theta^\mu_\mu(0)^\dagger|0\ra,
\eeq
where $\theta_{\mu\nu}$ is the improved QCD energy-momentum tensor 
(neglecting heavy quarks) whose anomalous trace reads, in standard 
notations:
\bea
\theta^\mu_\mu(x)&=&\frac{1}{4}\beta(\alpha_s)G^2+(1+\gamma_m(\alpha_s))
\sum_{u,d,s} m_i\bar \psi_i\psi_i.
\eea
Its leading-order perturbative and non-perturbative
expressions in $\alpha_s$ have been obtained by the authors of \cite{NSVZ}.
To two-loop accuracy 
in the $\overline{MS}$ scheme, its perturbative
expression has been obtained by \cite{KATAEV}, 
while the radiative
correction to the gluon condensate has been derived in
\cite{BAGAN}. Using a simplified
version of the
notation in Eq. (5):
\beq
\psi_s(q^2)=\sum_{D=0,4,...}C_D\la O_D\ra,
\eeq
 one obtains for three flavours and by normalizing the result with 
$(\beta(\alpha_s)/\alpha_s)^2$:
\bea
C_0&=&-2\asr^2 (-q^2)^2      
\log{-\frac{q^2}{\nu^2}}\aga 1+\frac{59}{4}\asr
+\frac{\beta_1}{2}\asr\log{-\frac{q^2}{\nu^2}}\adr\nnb\\
C_4\la O_4\ra&=&4\als\aga 1+\frac{49}{12}\asr
+\frac{\beta_1}{2}\asr\lnu\adr\gg \nnb\\
C_6\la O_6\ra&=& 2\as\aga 1-\frac{29}{4}\als \lnu\adr \ggg\nnb\\
C_8\la O_8\ra&=& 14 \la \ga\als f_{abc}G^a_{\mu\rho}G^{b\rho}_\nu\dr^2\ra
-\la \ga\als f_{abc}G^a_{\mu\nu}G^{b}_{\rho\lambda}\dr^2\ra .
\eea
From its asymptotic behaviour ($q^4\log$), we can write a 
twice-subtracted dispersion relation for $\psi_s(q^2)$:
\beq
\psi_s(q^2)=\psi_s(0)+ q^2\psi'_s(0)+
q^4\int_0^\infty \frac{dt}{t^2(t-q^2-i\epsilon)}
\frac{1}{\pi}\mbox{Im} \psi_s(t),
\eeq
where the subtraction constant $\psi_s(0)$ is known from the
LET to be \cite{NSVZ}:
\beq
\psi_s(0)=-16\frac{\beta_1}{\pi}\gg .
\eeq
\subsection{The gluonium sum rules}
Using standard QSSR technology, one can derive, from the previous expression
of the two-point correlator, the unsubtracted Laplace sum rule (USR):
\bea
{\cal L}_s^{(0)}&\equiv&\int_{t_\leq}^{\infty} {dt}~\mbox{exp}(-t\tau)
~\frac{1}{\pi}~\mbox{Im} \psi_s(t)\nnb\\
&=&\frac{4}{\pi^2}\beta^2(\alpha_s)
\tau^{-3}\Bigg{[}\Bigg{[} 1+\frac{1}{4}\as\ga
59+2\beta_1(1-2\gamma_E)
-4\frac{\beta_2}{\beta_1}\log{-\log({\tau\Lambda^2)}}\dr\Bigg{]}\nnb\\ 
&&-\pi\frac{\beta_1}{2}\gg\tau^2+
\frac{\pi}{2\alpha_s}\ggg\tau^3+\frac{2\pi^3}{\alpha_s}\gg^2\tau^4\Bigg{]}\nnb\\
{\cal L}_s^{(1)}&\equiv& -\frac{d}{d\tau}{\cal L}_s^{(0)},\nnb\\
{\cal L}_s^{(2)}&\equiv& -\frac{d}{d\tau}{\cal L}_s^{(1)}
\eea
and the subtracted sum rule (SSR), which depends crucially on
the subtraction constant $\psi_s(0)$:
\bea
{\cal L}_s^{(-1)}&\equiv&\int_{t_\leq}^{\infty} \frac{dt}{t}~\mbox{exp}(-t\tau)
~\frac{1}{\pi}~\mbox{Im} \psi_s(t)\nnb\\
&=&\frac{2}{\pi^2}\beta^2(\alpha_s)\tau^{-2}\Bigg{[} 1+\frac{1}{4}\as\ga
59+4\beta_1(1-\gamma_E)
-4\frac{\beta_2}{\beta_1}\log{-\log{(\tau\Lambda^2)}}\dr\Bigg{]}\nnb\\ 
&&+\psi_s(0)
-{4}\frac{\beta^2(\alpha_s)}{\alpha_s}\Bigg{[}\gg\Bigg{[} 1+\as\ga \frac{49}{12}
-\frac{\gamma_E}{2}\beta_1\dr\Bigg{]}\nnb\\
&&+\frac{\tau}{2\pi}\ggg+\pi\gg^2\tau^2\Bigg{]},
\eea
where:
\beq
\as\equiv \frac{1}{\beta_1\log{\sqrt{\tau}\Lambda}}.
\eeq
Throught this paper, we shall use for three active flavours
\cite{BETHKE}:
\beq
\Lambda= (375\pm 125)~\mbox{MeV}.
\eeq
$\b$ One can notice that the perturbative correction is large for
the Laplace sum rules ${\cal L}_s^{(n)}$, but it tends to cancel in the
ratio of moments. As discussed earlier, 
we estimate the higher-order unknown perturbative corrections
using a geometrical
growth of the QCD series. Then, we
expect that the ``effective"
$\alpha_s^2$ correction including $\log\log$ terms
normalized to the lowest-order perturbative graph is here
about:
\beq
\delta_{\alpha}^{(4)}\approx \pm 2\times 18 \as^2~,
\eeq
which we consider as the error
due to the truncation of the perturbative QCD series
(last known term of the series), which should include in it some
(eventual) effects of the 1/$Q^2$-term induced by the resuumation of this
series. The factor 2 is a conservative estimate of the errors.\\
$\b$ We shall see later on, that the non-perturbative contributions
 including the corresponding $\alpha_s$ corrections are small
at the scale where the sum rules are optimized, such that the dominant
errors from the non-perturbative effects come from the
values of the condensates.\\
$\b$ One should also notice that it can be inconsistent to work here
in a pure Yang-Mills, as
our QCD parameters ($\alpha_s$ and gluon condensates)
have been extracted from the data in 
the presence of
quarks. Therefore, one should be careful in comparing the results in the
present paper with e.g. the lattice ones.\\
$\b$ The USR ${\cal L}_s^{(0)}$ and the SSR ${\cal L}_s^{(-1)}$ 
have been used in NV. The authors of Ref. \cite{NSVZ} 
have also used the SSR in order to set 
the sum rule scale in the scalar gluonium channel, leading to the conclusion 
that 
the scalar gluonium energy scale should be much larger
(or $\tau$ much smaller) than the one of ordinary hadrons. We
shall see later on that the SSR ${\cal L}_s^{(-1)}$
satisfies this criteria, due mainly to the
relative importance of the $\psi_s(0)$ contribution. However,
this conclusion is not universal, as it
does not necessarily apply to the
USR, where $\tau$ can be slightly larger (or the sum rule energy
scale slightly lower) than in the SSR. A simple inspection of these different
sum rules indicates that ${\cal L}_s^{(0)}$ is the only sum rule that can be 
sensitive to a low-mass resonance below 1 GeV, as it is a low moment in $t$ and it 
can stabilize at large value of $\tau$. Higher
moments in $t$, ${\cal L}_s^{(1,2)}$,are,  on the contrary, sensitive to 
a higher mass resonance. 
As discussed before, ${\cal L}_s^{(-1)}$ is quite particular since, 
though it is the
lowest moment, which a priori can be more sensitive to the lowest meson mass,
its stability is pushed to higher energies because of the relatively
large value of $\psi_s(0)$. As a biproduct, ${\cal L}_s^{(-1)}$
becomes less sensitive to a low-mass resonance than  ${\cal L}_s^{(0)}$.
\subsection{Mass of the gluonium $G$}
In order to study the properties of the gluonium $G$ expected from other approaches
to be in the range of 1.4 to 2 GeV, we have to work with 
sum rules that are more sensitive to the high-energy region than 
${\cal L}_s^{(0)}$. 
Using the positivity of the spectral functions, an upper
bound on the gluonium mass squared can be obtained from the minimum (or
inflexion point) of the ratios \footnote{We have also studied the ratio 
${\cal R}_{10}$, but found no $\tau$ stability there.}:
\beq
{\cal R}_{21}\equiv \frac{{\cal L}_s^{(2)}}{{\cal L}_s^{(1)}},
\eeq
giving (see Fig. 2):
\beq
M_G\leq (2.16\pm 0.16\pm 0.14\pm 0.05)~\mbox{GeV},
\eeq
where the errors come respectively from the value of $\Lambda$, from
the estimated
unknown higher-order terms and from the gluon condensate. At such
a small value of $\tau$, where the sum rule is optimized, we expect that
high-dimension terms including instanton effects are highly 
suppressed. Combining these
errors in quadrature, we obtain:
\beq
M_G\leq (2.16\pm 0.22)~\mbox{GeV},
\eeq

\begin{figure}[H]
\begin{center}
\includegraphics[width=6cm]{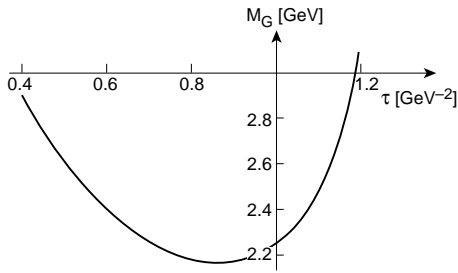}
\caption{ $\tau$ behaviour of the ratio of moments ${\cal R}^{1/2}_{21}$. 
Its minimum 
corresponds to an upper bound for the scalar gluonium mass. }
\end{center}
\end{figure}

\nin
An estimate of the mass squared can be obtained using the duality ansatz 
parametrization of the spectral functions, which leads to the FESR-like ratios
\beq
{\cal R}_{21}^c\equiv  \frac{\int_{t_\leq}^{t_c}{dt~}{t^2}~\mbox{exp}(-t\tau)
~\frac{1}{\pi}~\mbox{Im} \psi_G(t)~}
{\int_{t_\leq}^{t_c} {dt}~t~\mbox{exp}(-t\tau)
~\frac{1}{\pi}~\mbox{Im} \psi_G(t)}\simeq M^2_G,
\eeq
where $t_c$ is the QCD continuum threshold. Neglecting in the analysis the
eventual contribution of the light $\sigma_B$ of a mass
below 1 GeV, which is justified by the higher power of mass
suppression in the present sum rule, we show in Fig. 3a, b the
stabilities in $\tau$ (here it is
an inflexion point) and in
$t_c$ (here it is a minimum) of the estimate.

\begin{figure}[H]
\begin{center}
\includegraphics[width=10cm]{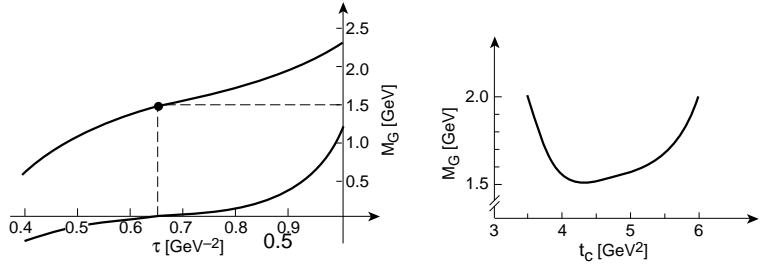}
\caption{ a): $\tau$ behaviour of the estimated value
of the scalar gluonium mass for $t_c=4.5$ GeV$^2$. The optimal value is
at the inflexion point corresponding to the zero of its 
second derivative in $\tau$.
b): $t_c$ behaviour  of $M_G$. }
\end{center}
\end{figure}

\nin
 Then,
we deduce the optimal result:
\beq
M_G \simeq (1.50\pm 0.10\pm 0.15\pm 0.05)~\mbox{GeV},
\eeq
where the errors are due to $\Lambda$, the estimated
unknown higher-order terms and the gluon condensate. Combining these
errors in quadrature, we obtain:
\beq
M_G\simeq (1.50\pm 0.19)~\mbox{GeV},
\eeq
in agreement with the most recent sum rule result in \cite{BAGAN}.
Within the sum rule approach, one
can approximately identify the value of $t_c$ as the mass squared of the next
radial excitation:
\beq
M_{G'}\approx \sqrt{t_c} \simeq (2.0\sim 2.1)~\mbox{GeV},
\eeq
where it can be noticed that, unlike the usual hadrons, the splitting 
between the 
lowest ground states and the radial excitations are relatively small 
($\Delta M_G/M_G\approx 30\%$ compared with 
$\Delta M_{\rho}/M_{\rho}\geq 70\%$).
One can compare our value of $M_G$ with the theoretical
estimates (lattice calculations \cite{SHARPE}--\cite{WEIN}, QCD inequalities
\cite{WEST}, ...)
and with the GAMS \cite{GAMS} and Crystal Barrel \cite{BARREL} data. 
\subsection{Decay constant of the G(1.5)}
The decay constant of the G(1.5) can be introduced via:
\beq
\la 0|4\theta^\mu_\mu|G\ra=\sqrt{2}f_{G}M^2_{G},
\eeq 
where $f_G$ is the decay constant analogue of $f_\pi$.
One can either estimate $f_G$, by using the previous
${\cal L}^{(1,2)}_s$ sum rules, or using the SSR ${\cal L}_s^{(-1)}$,
as in NV, which we shall discuss later on. Using the previous sum rules, for
instance ${\cal L}^{(1)}_s$, we obtain (see Fig. 4):
\beq
f_G\simeq (390\pm 98\pm 65\pm 39\pm 76)~\mbox{MeV},
\eeq 
where the errors come from $\Lambda$, the estimated
unknown higher-order terms, the gluon condensate and the value of $t_c$. 
Combining these errors in quadrature, we obtain:
\beq
f_G\simeq (390\pm 145)~\mbox{MeV},
\eeq 
which we shall use later on. However, one should remark that the errors in the present
determinations are relatively large compared with the typical 10-20$\%$ accuracy in
the QSSR estimate of the
corresponding quantity in the ordinary hadron sector.

\begin{figure}[H]
\begin{center}
\includegraphics[width=6cm]{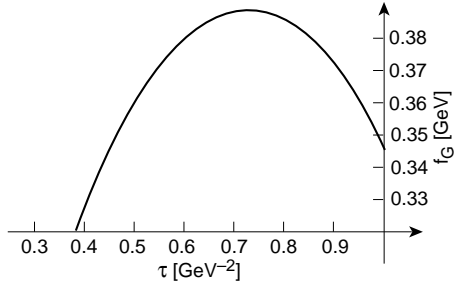}
\caption{$\tau$-behaviour of the decay constant $f_G$ from ${\cal L}^{(1)}_s$.}
\end{center}
\end{figure}

\nin
\subsection{Mass and decay constant of the three-gluon bound state}
Recall that a QSSR analysis of the two-point correlator associated to
the scalar three-gluon local current to leading order \cite{LATORRE,SNB}:
\beq
J_{3G}=g^3f_{abc}G^aG^bG^c
\eeq
leads to the mass prediction:
\beq
M_{3G}\simeq 3.1~\mbox{GeV}~~~~~~~~\sqrt{t_c}\simeq 3.4~\mbox{GeV},
\eeq
and to the value of the decay constant of:
\beq
f_{3G}\simeq 62~\mbox{MeV},
\eeq
which is relatively high compared with the mass of the gluonium built from
the two-gluon current, and which makes the mass-mixing between these two gluonia
states tiny \cite{LATORRE,SNB}:
\beq
\theta_3\approx 4^\circ~,
\eeq 
where the accuracy for the above predictions are about 10-20$\%$. This state
might be produced in the radiative decay of heavy quarkonia, when phase space
permits, while its experimental search can be done by measuring some typical
gluonia decays into the pairs $\eta'\eta'$ or $\eta\eta'$.  
\subsection{Decay constants of the $\sigma_B$ and $\sigma'_B$ mesons}
As discussed previously and in NV, one can expect that the low moments
${\cal L}_s^{(-1,0)}$ are sensitive to the low-mass resonances whose
effects can have been missed in the previous analysis of 
${\cal L}_s^{(1,2)}$, and presumably in the
lattice calculations within a one-resonance parametrization.
In the following, we shall therefore test the ``gluonium" nature of the
broad low-mass states $\sigma_B$ and $\sigma'_B$, where the former, which 
has
a mass in the range 0.5 to 1 GeV, is the
one seen in $\gamma\gamma$ and $\pi\pi$ scattering experiments, and 
expected
from the linear $\sigma$ model \cite{LANDUA,ANISO}, while we identify the
second as its radial excitation with a mass close to that of the observed
state at 1.37 GeV.
We use a three-resonance ($\sigma_B$, $\sigma'_B$ and $G(1.5)$)
parametrization of the spectral function \footnote{Unfortunately, 
we cannot fix 
simultaneously the masses and decay constants of the 
$\sigma_B$ and $\sigma'_B$ with the two
sum rules.}. 
We introduce the
previous values of the $G(1.5)$ parameters and the corresponding
value of $t_c\simeq 4.5$ GeV$^2$. Using as input $M_{\sigma'_B}\approx
1.37$ GeV and $M_{\sigma_B}\approx
(0.5\sim 1)$ GeV,
we extract from the two sum rules the decay
constants of the $\sigma_B$ and $\sigma'_B$ within
the $\tau$-stability criteria. We show in  Fig. 5 the $\tau$
behaviour of:
\beq
\tilde f_2\equiv f_{\sigma_B}\ga 1+\rho ~\exp{(M^2_{\sigma'_B}-
M^2_{\sigma_B})\tau}\dr^{1/2},
\eeq
and of:
\beq
\tilde f_4\equiv f_{\sigma_B}\ga 1+\rho\frac{M^2_{\sigma'_B}}{
M^2_{\sigma_B}} ~\exp{(M^2_{\sigma'_B}-
M^2_{\sigma_B})\tau}\dr^{1/2},
\eeq
where:  
\beq
\rho\equiv \frac{M^2_{\sigma'_B}f_{\sigma'_B}}{M^2_{\sigma_B}
f_{\sigma_B}}.
\eeq 

\begin{figure}[H]
\begin{center}
\includegraphics[width=10cm]{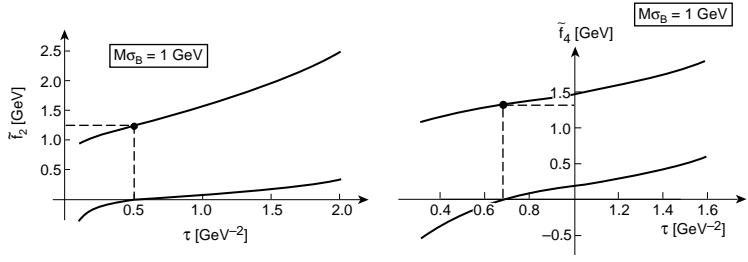}
\caption{ $\tau$ behaviour of the decay constants 
$\tilde f_{2}$ from ${\cal L}^{(-1)}_s$ and
$\tilde f_{4}$ from ${\cal L}^{(0)}_s$ }
\end{center}
\end{figure}

\nin
Then, we deduce:
\bea
f_{\sigma_B}[\mbox{GeV}]\approx &&1.0~~~~~
~~~~~ f_{\sigma'_B}[\mbox{GeV}]\approx 0.6~~~~~\mbox{for}~~~~
M_{\sigma_B}[\mbox{GeV}]\approx  
1.0\nnb\\
&&1.4~~~~~~~~~~~~~~~~~~~~~~~~~~~0.7~~~~~~~~~~~~~~~~~~~~~~~~~~~~~~~
0.75\nnb\\
&&1.9~~~~~~~~~~~~~~~~~~~~~~~~~~~0.5~~~~~~~~~~~~~~~~~~~~~~~~~~~~~~~0.5~,
\eea
which indicates the necessity to have ``low-mass gluonia"
(which,  as we shall see later on,
 couples strongly to $\pi\pi$) 
for the consistency
of the sum rules approach \footnote{The existence of these
states is also needed in the analysis of the
correlator, by combining
the LET result for $\psi_s(0)$
with dispersive techniques \cite{SHIFMAN}.}, 
although their effects are negligible in the
high-moments analysis. 
The previous quantities will be useful
later on for studying the $\sigma_B$ and $\sigma'_B$ decays.
\section{Decay widths of the $\sigma_B$, $\sigma'_B$ and $G$}
\subsection{$\sigma_B$ and $\sigma'_B$ couplings to $\pi\pi$}
At this point, we use vertex sum rules to obtain further constraints. 
We consider the vertex (Fig. 6a):
\beq
V(q^2)=\la\pi_1|\theta^\mu_\mu|\pi_2\ra,~~~~~q=p_1-p_2~,
\eeq
where:
\beq
V(0)=2m^2_\pi~. 
\eeq

\begin{figure}[H]
\begin{center}
\includegraphics[width=8cm]{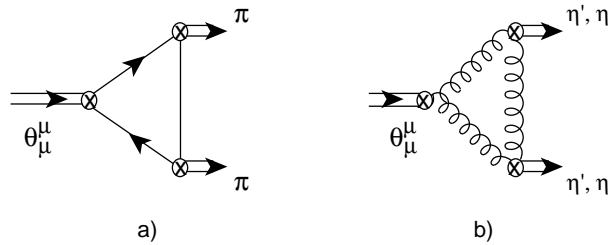}
\caption{Vertex controlling the gluonium couplings to: a) $\pi\pi$; b) 
$\eta(\eta')-\eta(\eta')$. }
\end{center}
\end{figure}

\nin
In the chiral limit $(m^2_\pi \simeq 0)$, 
the vertex obeys the dispersion relation:
\beq
V(q^2)=\int_0^\infty \frac{dt}{t-q^2-i\epsilon}
~\frac{1}{\pi}\mbox{Im} V(t),
\eeq
which gives, by saturating with the three resonances $\sigma_B$,
$\sigma'_B$ and $G$:
\beq
\frac{1}{4}\sum_{S=\sigma_B,\sigma'_B,G}g_{S\pi\pi}\sqrt{2}f_S \simeq 0.
\eeq
Using the fact that $ V^{\prime} (0)=1$ \cite{NSVZ2}
from a generalization of the Goldberger-Treiman relation
using soft-pion techniques, one obtains a second sum rule:
\beq
\frac{1}{4}\sum_{i=\sigma_B,\sigma'_B,G}g_{S\pi\pi}\sqrt{2}f_S/M^2_S=1.
\eeq
Identifying the $G$ with the $G(1.5\sim 1.6)$ at GAMS and the Crystal Barrel, 
we can neglect then its coupling to $\pi\pi$. Therefore, we can deduce:
\bea
g_{\sigma_B\pi\pi}&\approx& \frac{4}{\sqrt{2}f_{\sigma_B}}~\frac{1}
{\ga 1- 
M^2_{\sigma_B}/M^2_{\sigma'_B}\dr}\nnb\\
g_{\sigma'_B\pi\pi}&\approx&g_{\sigma_B\pi\pi}
\ga \frac{f_{\sigma_B}}{f_{\sigma'_B}}\dr,
\eea
Fixing for definiteness $M_{\sigma'_B}
\simeq 1.37$ GeV, one can deduce the width into $\pi\pi~ (\pi^+\pi^-,~2\pi^0)$:
\bea
\Gamma(\sigma_B\rar\pi\pi)[\mbox{GeV}]&\approx &0.7~~~~~
~~~~~ \Gamma(\sigma'_B\rar\pi\pi)[\mbox{GeV}]\approx~~ 1.3~~~~~
\mbox{for}~~~~
M_{\sigma_B}[\mbox{GeV}]\approx  
1.0\nnb\\
&&0.2~~~~~~~~~~~~~~~~~~~~~~~~~~~~~~~~~~~~~~~~0.5
~~~~~~~~~~~~~~~~~~~~~~~~~~~~~~~0.75\nnb\\
&&0.1~~~~~~~~~~~~~~~~~~~~~~~~~~~~~~~~~~~~~~~~0.9
~~~~~~~~~~~~~~~~~~~~~~~~~~~~~~~0.5~,
\eea
where
\beq
\Gamma(\sigma_B\rar\pi\pi)=\frac{|g_{\sigma_B\pi\pi}|^2}
{16\pi M_{\sigma_B}}\ga{1-\frac{4m^2_\pi}{M^2_{\sigma_B}}}\dr^{1/2}.
\eeq
We have repeated the derivation of $f_{\sigma_B}$ by taking into account 
finite-width corrections. This leads to an increase of $f_{\sigma_B}$, 
which is compensated
by the propagator effect in the estimate of $g_{\sigma_B\pi\pi}$, i.e. 
the
result obtained from the vertex sum rule remains almost unchanged. It is 
interesting
to see from the sum rule that a very light $\sigma_B$ around $500$ MeV 
cannot 
be broad, which seems not to be
favoured by the present data. Improvements of the data analysis are 
required 
for refining the mass
measurement of the $\sigma_B$. Our result indicates 
the presence of gluons inside the wave 
functions of the broad resonance below 1 GeV and the $\sigma'(1.37)$, which 
can decay copiously into $\pi\pi$ \footnote{The decays of the physically 
observed
states will be discussed later on.}. 
\subsection{{\it G}(1.5) coupling to $\eta$$\eta'$}
In order to compute these couplings, we consider the three-point function
(Fig. 6b):
\beq
\tilde{V}_{\mu\nu}(q_1,q_2)\equiv \int d^4x_1~d^4x_2~e^{i(q_1x_1+q_2x_2)}
~\la 0| {\cal T} Q(x_1)Q(x_2)\theta_{\mu\nu}(0) |0\ra~,
\eeq
where $\theta_{\mu\nu}$ is the energy momentum tensor of QCD with three 
light quarks, while $Q(x)$ is the topological charge density defined 
previously.
Using the low-energy theorem:
\beq
\la \eta_1|\theta^\mu_\mu|\eta_1\ra = 2M^2_{\eta_1},
\eeq
where $\eta_1$ is the unmixed $U(1)$ singlet state of mass
$M_{\eta_1}\simeq $ 0.76 GeV \cite{WITTEN}, and
writing the dispersion relation for the vertex, one obtains the NV 
sum rule:
\beq
\frac{1}{4}\sum_{S\equiv\sigma_B,\sigma'_B,G}g_{S\eta_1\eta_1}\sqrt{2}f_S=
2M^2_{\eta_1},
\eeq
which implies, for $M_{\sigma_B}\simeq (0.75\sim 1)$ GeV,
and by assuming a $G$-dominance of the
vertex sum rule:
\beq
g_{G\eta_1\eta_1}\approx (1.2\sim 1.7)~\mbox{GeV}.
\eeq
Introducing the ``physical" $\eta'$ and $\eta$ through:
\bea 
\eta'\sim \cos\theta_P \eta_1-\sin\theta_P \eta_8\nnb\\
\eta\sim \sin\theta_P \eta_1+\cos\theta_P \eta_8,
\eea
where \cite{PDG,GILMAN}
\beq
\theta_P\simeq -(18\pm 2)^\circ
\eeq
is the pseudoscalar mixing angle,
one obtains:
\beq
\Gamma (G\rar\eta\eta')\simeq (5\sim 10)~ \mbox{MeV}.
\eeq
The previous scheme is also known to predict (see NV and \cite{GERS}):
\beq
r\equiv \Gamma_{G\eta\eta}/\Gamma_{G\eta\eta'}\simeq 0.22,~~~~~~~~~~~~
g_{G\eta\eta}\simeq \sin\theta_Pg_{G\eta\eta'}~~~\lrar \Gamma_{G\eta\eta}
\simeq (1.1\sim 2.2)~\mbox{MeV}.
\eeq
compared with the GAMS data \cite{GAMS} $r\simeq 0.34\pm 0.13$. This result,
can then suggest that the $G(1.6)$ seen by the GAMS group is a pure gluonium, which
is not the case of the particle seen by Crystal Barrel \cite{BARREL}.
\subsection{$\sigma'_B(1.37)$ and {\it G}(1.5) couplings 
to $4\pi$ through $\sigma_B\sigma_B$}
Within our scheme, we expect that the $4\pi$ are mainly initiated 
from the decay of pairs of $\sigma_B$. A LET
analogous to that in previous cases, can be written for 
the 
estimate 
of $g_{G\sigma_B\sigma_B}$. Using
\beq
\la \sigma_B|\theta^\mu_\mu|\sigma_B\ra = 2M^2_{\sigma_B}
\eeq
and writing the dispersion relation for the vertex, one obtains the sum 
rule:
\beq
\frac{1}{4}\sum_{i=\sigma_B,\sigma'_B,G}g_{S\sigma_B\sigma_B}\sqrt{2}f_S=
2M^2_{\sigma_B}.
\eeq
We identify the $\sigma'_B$ with the observed $f_0(1.37)$. Using
the observed value of the width into $4\pi$ 
\cite{PDG,GASPERO} and extracting the $S$-wave part, which we assume to
be initiated from $2\sigma_B$, we deduce \footnote{We have taken the 
largest range
deduced from the different branching ratios given by PDG \cite{PDG}. 
We have used
$\Gamma(\sigma_B\rar {\mbox all}\simeq (150\sim 500)$ MeV.}:
\beq
\Gamma(\sigma'_B\rar (4\pi)_S) \simeq (46\sim 316)~\mbox{MeV}~\lrar~
 g_{\sigma'_B\sigma_B\sigma_B}\simeq (2.3\sim 5.9)~\mbox{GeV},
\eeq
where we have taken finite width corrections 
from the Breit-Wigner
parametrization of the $\sigma_B$ resonance. We have
used $M_{\sigma_B}\simeq 1$ GeV and $\Gamma(\sigma_B\rar\pi\pi)\simeq 0.7$ 
GeV.
Neglecting, to a first approximation, the $\sigma_B$ contribution
to the sum rule, we can deduce:
\beq
g_{G\sigma_B\sigma_B}\approx (2.7\sim 4.3)~\mbox{GeV},
\eeq
which, despite the large error,
is larger than the $G\eta_1\eta_1$ coupling. It can indicate that
the decay of a pure gluonium state can be dominated (phase space
permitting) by the $4\pi$ branching
ratio initiated from the pair of $\sigma$-mesons, as already emphasized
by NV. Using the previous values
and the corresponding decay
constant and width, one can deduce, using a Breit-Wigner form:
\beq
\Gamma(G\rar\sigma_B\sigma_B\rar 4\pi)\simeq (60\sim 138)~\mbox{MeV}.
\eeq
This feature seems to be satisfied
by the states seen by GAMS and the Crystal Barrel. Our approach shows 
the consistency in
interpreting the $G(1.6)$ seen at GAMS as an ``almost" pure gluonium state
(ratio of the $\eta\eta'$ versus the $\eta\eta$ widths), 
while the
state seen by the Crystal Barrel, though having a gluon component 
in its wave function,
cannot be a pure gluonium because of its prominent
decays into $\eta\eta$ and $\pi^+\pi^-$. We shall see later on
that the Crystal Barrel state
can be better explained from a quarkonium-gluonium mixing.
\subsection{$\sigma_B$, $\sigma'_B$ and $G$ couplings to $\gamma\gamma$}
The two-photon widths of the $\sigma,~\sigma'$ and $G$
 can be obtained by identifying the 
Euler-Heisenberg effective Lagrangian (Fig. 7a) \cite{NSVZ2}:
\bea
{\cal L}_{\gamma g}&=&\frac{\alpha\alpha_sQ^2_q}{180m^2_q}\lb 
28F_{\mu\nu}F_{\nu\lambda}G_{\lambda\sigma}G_{\sigma\mu}
+14F_{\mu\nu}G_{\nu\lambda}F_{\lambda\sigma}G_{\sigma\mu}\nnb\\
&-&F_{\mu\nu}G_{\mu\nu}F_{\alpha\beta}G_{\alpha\beta}-
F_{\mu\nu}F_{\mu\nu}G_{\alpha\beta}G_{\alpha\beta}\rb,
\eea
where $Q_q$ is the quark charge in units of $e$, $-\beta_1=9/2$ for three
 flavours, and $m_q$ is the ``constituent" quark mass, which we shall
take to be
\beq
m_u\simeq m_d \simeq M_\rho/2,~~~~~~~~m_s\simeq M_\Phi/2~,
\eeq
with the scalar-$\gamma\gamma$ Lagrangian
\beq
{\cal L}_{S\gamma\gamma}=g_{S\gamma\gamma}\sigma_B(x)
F^{(1)}_{\mu\nu}F^{(2)}_{\mu\nu}.
\eeq

\begin{figure}[H]
\begin{center}
\includegraphics[width=6cm]{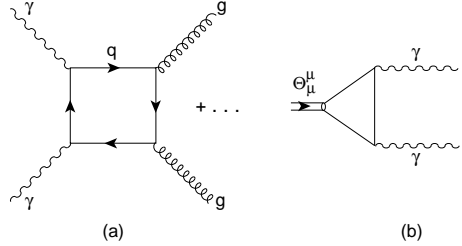}
\caption{Vertex controlling the gluonium couplings 
to $J/\psi(\gamma)\gamma$: a) box diagram; b) 
anomaly diagram. }
\end{center}
\end{figure}

\nin
This leads to the sum rule:
\beq
{g_{S\gamma\gamma}}\simeq
\frac{\alpha}{60}\sqrt{2}{f_{\sigma_B}M^2_{\sigma_B}}
\ga\frac{\pi}{-\beta_1}\dr\sum_{q\equiv u,d,s}{Q^2_q/m_q^4},
\eeq
from which we deduce the couplings
\footnote{Here and in the following, we shall use
$M_{\sigma_B}\approx (0.75\sim 1.0)$ GeV.}:
\beq
g_{\sigma_B\gamma\gamma}\approx (0.4\sim 0.7)\alpha~\mbox{GeV}^{-1}
~~~~~~~~~~~~~~~~~~~
g_{\sigma'_B\gamma\gamma}\approx (0.5\sim 0.6)\alpha~\mbox{GeV}^{-1}
~~~~~~~~~~~~~~~~~~~
g_{G\gamma\gamma}\approx (0.6\pm 0.2)\alpha~\mbox{GeV}^{-1}.
\eeq
Using the corresponding decay width:
\beq
\Gamma(S\rar\gamma\gamma)=\frac{|g_{S\gamma\gamma}|^2}
{16\pi}M^3_{S},
\eeq
one obtains:
\beq
\Gamma(\sigma_B\rar\gamma\gamma)\approx (0.2\sim 0.3)~\mbox{ keV}~~~~~~~
\Gamma(\sigma'_B\rar\gamma\gamma)\approx (0.7\sim 1.0)~\mbox{ keV}~~~~~~~
\Gamma(G\rar\gamma\gamma)\approx (1.0\pm 0.8)~\mbox{ keV},
\eeq
in agreement with the NV results, but smaller (as expected
from general grounds) than the  
well-known quarkonia widths:
\beq
\Gamma(\eta'\rar\gamma\gamma)\simeq 4.2~\mbox{ keV}~~~~~~~
\Gamma(f_2\rar\gamma\gamma)\simeq 2.6~\mbox{ keV}.
\eeq
Alternatively,  one can use the trace anomaly (Fig. 7b):
\beq
\la 0|\theta^\mu_\mu|\gamma_1\gamma_2\ra=\la 0|\frac{1}{4}\beta(\alpha_s)G^2+
\frac{\alpha R}{3\pi}F^{\mu\nu}_1F^{\mu\nu}_2|\gamma_1\gamma_2\ra,
\eeq
where $F^{\mu\nu}$ is the photon field strength and $R\equiv 3\sum Q^2_i$. 
Using the 
fact that the RHS is ${\cal O}(k^2)$ implies the sum rule \cite{ELLIS,NSVZ}:
\beq
\la 0|\frac{1}{4}\beta(\alpha_s)G^2|\gamma_1\gamma_2\ra =-
\la 0|\frac{\alpha R}{3\pi}F^{\mu\nu}_1F^{\mu\nu}_2|\gamma_1\gamma_2\ra,
\eeq
from which one can deduce the coupling:
\beq
\frac{\sqrt{2}}{4}\sum_{S\equiv\sigma_B,\sigma'_B,G}f_Sg_{S\gamma\gamma}\simeq 
\frac{\alpha R}{3\pi}.
\eeq
It is easy to check that the previous values of the couplings also
satisfy the trace anomaly sum rule.
 However, the result for the $\sigma\gamma\gamma$ coupling
is much smaller than the ones in \cite{SHARPE2,CHAN} obtained from a single
resonance saturation of the trace anomaly, because (as we have seen previously)
the one resonance saturation is not a good approximation, while 
the value of $f_{\sigma_B}$
used in these papers is much smaller than the one obtained here. 
\subsection{$\sigma_B$,
$\sigma'_B$ and $G$ productions from radiative $J/\psi$ decays}
As in \cite{NSVZ2}, one can estimate this process, using dipersion relation
techniques, by saturating the spectral function by the $J/\psi$ plus a
continuum. The glue part of the amplitude can be converted into a physical
non-perturbative
matrix element $\la 0|\alpha_s G^2|S\ra$ known through the decay constant $f_S$
estimated from QSSR.
By assuming that the continuum is small, one obtains:
\beq
\Gamma(J/\psi\rar\gamma S)\simeq \frac{\alpha^3\pi}{\beta_1^2 656100}
\ga\frac{M_{J\psi}}{M_c}\dr^4\ga\frac{M_{S}}{M_c}\dr^4 
\frac{\ga 1-M^2_S/M^2_{J/\psi}\dr^3}{\Gamma(J/\psi\rar e^+e^-)}f^2_S,
\eeq
where $M_c\simeq 1.5$ GeV is the charm constituent quark mass. 
We use $-\beta_1=7/2$ for six flavours. This leads to the rough
estimates:
\bea
B(J/\psi\rar\gamma\sigma_B)\times B(\sigma_B\rar~ \mbox{all})
&\approx& (0.4\sim 0.6)\times 10^{-3}\nnb\\ 
B(J/\psi\rar\gamma\sigma'_B)\times B(\sigma'_B\rar~ \mbox{all})
&\approx& (0.8\sim 1.0)\times 10^{-3}\nnb\\ 
B(J/\psi\rar\gamma G)\times B(G\rar~\mbox{all})&\approx&(0.5\sim 0.4)\times 10^{-3}.
\eea
These branching ratios can be compared with the observed 
$B(J/\psi\rar\gamma\eta')$
and $B(J/\psi\rar\gamma f_2)$ ones, which are respectively $4\times 10^{-3}$
and $1.6\times 10^{-3}$. The $\sigma_B$ could already have been
 produced, but might
have been confused with the $\pi\pi$
background. The ``pure gluonium"
$G$ production rate is relatively small,
contrary to the na\"{\i}ve expectation for a glueball production. In our
approach, this is due to the relatively small value of its decay constant,
which controls the non-perturbative dynamics. 
Its observation from this process should wait for the $\tau$CF machine. 
However, we do not exclude the possibility that a state resulting from a 
quarkonium-gluonium mixing may be produced at higher rates. 
From the previous results, one can also deduce the corresponding
stickiness defined in \cite{CHAN}.
\section{Properties of the scalar quarkonia}
\subsection{Mass and decay constant of the $S_2(\bar uu+\bar dd)$ 
quarkonium}
We consider this state as the $SU(2)$ partner of the $a_0(0.98)$ associated 
to the divergence of the charged vector current of current algebra:
\beq
\partial_\mu V^\mu(x)\equiv (m_u-m_d)\bar u(i\gamma_5)d.
\eeq
The mass and coupling of the $a_0$ have been studied within the QSSR 
\cite{SNB}:
\beq
M_{a_0}\simeq (1\sim 1.05)~\mbox{GeV}~~~~~f_{a_0}\simeq (0.5\sim 1.8)
~\mbox{MeV},
\eeq
where the small value of $f_{a_0}$ is due to the light current quark mass 
difference.
In our approach, due to the good realization of
the $SU(2)$ symmetry, the mass of the $S_2(\bar uu +\bar dd)$
bound state is expected to be degenerate with the one of the 
$a_0(\bar ud)$. 
The continuum threshold at which the previous parameters have been 
optimized can 
roughly indicate the mass of the next radial excitation, which is 
\cite{SNB}:
\beq
M_{S'_2}\approx \sqrt{t_c}\simeq (1.1\sim 1.4) ~\mbox{GeV}\approx M_{\pi'},
\eeq
which is about the $f_0(1.37)$ mass. An estimate of this mass using a model
with an infinite number of resonances leads also to the same result (see e.g.
S.G. Gorishny et al. in \cite{BECCHI}). 
\subsection{Couplings of the $S_2$ to $\pi^+\pi^-$, $K^+K^-$
and $\gamma\gamma$}
Using vertex sum rules, BN \cite{BRAMON2} obtain 
the $S_2$ coupling to pair of pions in the chiral limit:
\beq
g_{S_2\pi^+\pi^-}\simeq \frac{16\pi^3}{3\sqrt{3}}\la 
\bar uu\ra\tau e^{M^2_2\frac{\tau}{2}}\simeq 2.46~ \mbox{GeV},
\eeq
for the typical value of $\tau=1$ GeV$^{-2}$, in good agreement with the
one from the $SU(3)$ relation between the $S_2\rar\pi\pi$ and $a_0\rar\eta\pi$
widths and
with $g_{a_0K^+K^-}\simeq 2$ GeV from \cite{PAVER}, 
as intuitively expected. Therefore, we deduce with the same accuracy
as the one of the measured $a_0\rar\eta\pi$ width:
\beq
\Gamma(S_2\rar\pi^+\pi^-)\simeq  120 ~\mbox{MeV}.
\eeq
 Using $SU(3)$ symmetry, one can also 
expect:
\beq
g_{S_2 K^+K^-}\simeq  \frac{1}{2}g_{S_2\pi^+\pi^-}.
\eeq
The $\gamma\gamma$ width can also be obtained from the prediction in BN:
\beq
\Gamma(S_2\rar\gamma\gamma)\simeq \frac{25}{9}\Gamma(a_0(.98)\rar
\gamma\gamma)
\simeq 0.67~\mbox{keV},
\eeq
where 25/9 is the ratio of the $u$ and $d$ quark charges,
the last number coming from the data. This result is  
supported by the vertex sum rule analysis \cite{FOUR}.\\
The estimate of the $\gamma\gamma$ and hadronic
widths of the $S'_2$ is more uncertain. Using
the phenomenological observation that the coupling of the radial excitation
increases as the ratio of the decay constants $r\equiv
f_{S_2}/f_{S'_2}$, we expect:
\bea
\Gamma(S'_2\rar\gamma\gamma)&\approx& r^2\ga\frac{M_{S'_2}}{M_{S_2}}\dr^3
\Gamma(S_2\rar\gamma\gamma),\nnb\\
\Gamma(S'_2\rar\pi\pi)&\approx& r^2\ga\frac{M_{S_2}}{M_{S'_2}}\dr
\Gamma(S_2\rar\pi\pi),
\eea
which by taking $r\approx (M_{S'_2}/M_{S_2})^{(n=2)}$, like in the pion case
\cite{SNB} gives
\footnote{We estimate the error by assuming that $n=2\pm 1$}:
\beq
\Gamma(S'_2\rar\gamma\gamma)\approx (4\pm 2)~\mbox{keV},~~~~~~~~~
\Gamma(S'_2\rar\pi^+\pi^-)\approx (300\pm 150)~\mbox{MeV}.
\eeq
To a first approximation, we expect that the decay of the $S'_2$ into
$4\pi$ comes mainly from the pair of $\rho$ mesons, while the one from
$2\sigma_B$ (gluonia) is relatively suppressed as $\alpha_s^2$ 
using perturbative QCD arguments.
\subsection{Mass and decay constant of the $S_3(\bar ss)$ quarkonium}
In order to complete our discussions in the scalar sector, we compute 
from the 
sum
rules the mass and decay constant of the $S_3(\bar ss)$ state. 
In so doing, 
we work
with the two-point correlator:
\beq
\psi_{\bar ss}(q^2) \equiv i \int d^4x ~e^{iqx} \
\la 0\vert {\cal T}
J_{\bar ss}(x)
\ga J_{\bar ss}(0)\dr ^\dagger \vert 0 \ra ,
\eeq
where:
\beq
J_{\bar ss}(x)=m_s\bar ss,
\eeq
and we introduce the $S_3$ as:
\beq
\la 0|J_{\bar ss}|S_3\ra=\sqrt{2}f_{\bar ss}M^2_{\bar ss}. 
\eeq
We work with the Laplace transform sum rules:
\bea
{\cal F}_{\bar ss}(\tau)&\equiv& \int_{4m^2_s}^{\infty} dt \exp{(-t\tau)}
~\frac{1}{\pi}
\mbox{Im}\psi_3(t),\nnb\\
{\cal R}_{\bar ss}&\equiv&
 \frac{\int_{4m^2_s}^{\infty} dt~t~ \exp{(-t\tau)}~\frac{1}{\pi}
\mbox{Im}\psi_{\bar ss}(t)}{\int_{4m^2_s}^{\infty} dt \exp{(-t\tau)}~
\frac{1}{\pi}
\mbox{Im}\psi_{\bar ss}(t)}\simeq M^2_{\bar ss},\nnb\\
\frac{{\cal R}_{\bar ss}}{{\cal R}_{\bar us}}&\simeq& \frac{M^2_{\bar ss}}
{M^2_{K^*_0 (1.43)}}.
\eea
The QCD expressions of the sum rule have been
obtained in \cite{BECCHI} and are now known to three-loop accuracy (see
the compilation in \cite{JAMIN,CHET}).  
A much better stability is obtained at $\tau \simeq 0.4\sim 0.6$
GeV$^2$ by working with the double ratio of sum rules instead of the
 ratio. Using $\overline{m}_s(1$GeV)~= $(150\sim 190)$ MeV correlated 
to the values of $\Lambda$ \cite{SNM}, we deduce (see Fig. 8a,b):
\beq
M_{\bar ss}/M_{K^*_0 (1.43)}\simeq 1.03\pm 0.02
\lrar M_{\bar ss} \simeq (1474\pm 44)~\mbox{MeV}.
\eeq
This result confirms the earlier QSSR estimate in \cite{SNB}
\footnote{After obtaining this result, we learned
that a similar value has been obtained from
lattice calculations \cite{WEING}.}.
\begin{figure}[H]
\begin{center}
\includegraphics[width=10cm]{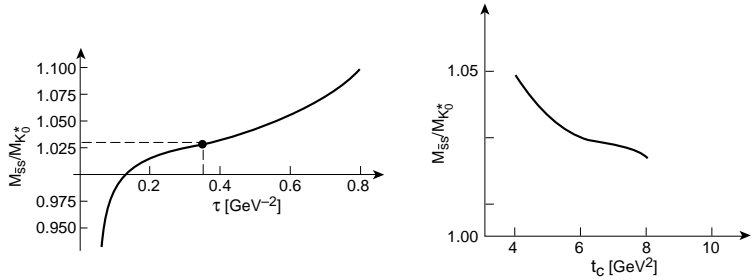}
\caption{$\tau$ and $t_c$ behaviours of $M_{\bar ss}/M_{K^*_0 (1.43)}$.}
\end{center}
\end{figure}

\nin
The result indicates the mass hierarchy:
\beq
M_{S_2\equiv \bar uu+\bar dd}
<M_{K^*_0\equiv \bar us}<M_{\bar ss}.
\eeq
The $SU(3)$ breaking obtained here is slightly larger than the na\"{\i}ve 
expectation
as, in addition to the strange-quark mass effect, the $\la \bar ss\ra$ 
condensate also plays
an important role in the splitting.
The sum rule
${\cal F}_{\bar ss}$ leads to the value of the decay constant (Fig. 9):
\beq
f_{\bar ss}= (43\pm 19)~\mbox{MeV}.
\eeq

\begin{figure}[H]
\begin{center}
\includegraphics[width=6cm]{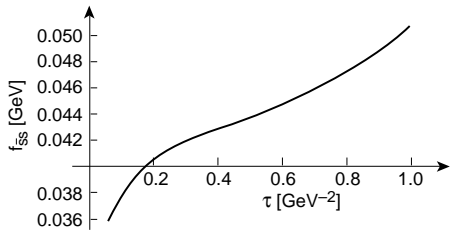}
\caption{$\tau$ behaviour of the decay constant $f_{\bar ss}$.}
\end{center}
\end{figure}

\nin
These previous estimates have been optimized at $t_c\simeq 6$ GeV$^2$. 
Therefore,
we expect that the radial excitation $S'_3$ will be in the range:
\beq
M_{S'_3}\approx (1.7\sim 2.4)~\mbox{GeV}, 
\eeq 
where the first number corresponds to the phenomenological
extrapolation $M_{S'_2}-M_{S_2}\simeq M_{S'_3}-M_{S_3}$, while the 
second value
is $\sqrt{t_c}$.
\subsection{Couplings of the $S_3$ to $K^+K^-$ and $\gamma\gamma$}
In so doing, we work with the vertex (Fig. 10):
\beq
V(p,q)=\int d^4x~d^4y~ e^{-ipx}e^{i(p+q)y}\la 0|{\cal T}J_{K^+}(x) J_{K^-}(y)
J_{\bar ss}(0)|0\ra~,
\eeq
and evaluate it at the symmetric point $p^2=q^2=(p+q)^2=-(Q^2\geq 0)$ as
in \cite{PAVER}.

\begin{figure}[H]
\begin{center}
\includegraphics[width=10cm]{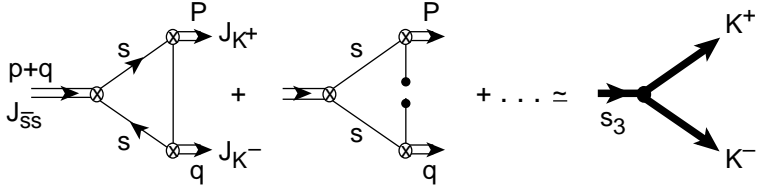}
\caption{Vertex sum rule for the $S_3K^+K^-$ coupling.}
\end{center}
\end{figure}

\nin
Then, we obtain, in $4-\epsilon$ dimensions:
\beq
V(-Q^2)=-m^2_s\aga -\frac{3}{2\pi^2}m_s\Gamma(\epsilon/2)\ga \frac{Q^2}
{\nu^2}\dr
^{-\epsilon/2}+\frac{1}{2Q^2}\la \bar uu-\bar ss\ra\adr .
\eeq
Its phenomenological expression can be approximated by:
\beq
V(-Q^2)=\frac{2f_K^2M^4_K}{\ga Q^2+M^2_K\dr^2} \frac{2f_{\bar ss}^2M^4_
{\bar ss}}{\ga Q^2+M^2_{\bar ss}\dr}
g_{S_3K^+K^-} .
\eeq
The Laplace transform of the previous equation leads to:
\beq
g_{S_3K^+K^-}\sqrt{2}M^2_{\bar ss}f_{\bar ss}\int_0^1xdx~
e^{-x(M^2_{\bar ss}-M^2_K)\tau}\simeq 4m_s\aga m_s-\frac{\pi^2}{3}
\tau\la \bar uu-2\bar ss\ra\adr .
\eeq
Using \cite{SNM}
 $\la\bar ss\ra/\la \bar uu\ra  \simeq 0.7$, $\la \bar uu\ra$(1 GeV) $\simeq 
-(229\mbox{MeV})^3$ and $m_s$ (1 GeV)$ ~\simeq$ (150$\sim$190) MeV, we obtain:
\beq
g_{S_3K^+K^-}\simeq (2.7\pm 0.5)~\mbox{GeV}~~~\mbox{and}~~~
g_{S_3\eta\eta}\simeq 0.9\sqrt{\frac{2}{3}}g_{S_3K^+K^-}.
\eeq
We can also predict:
\beq
g_{S_3\eta\eta'}\simeq \sin\theta_P g_{S_3\eta\eta}.
\eeq
This result can be compared with the one obtained previously, 
and with $g_{a_0K^+K^-}\simeq 2$ GeV from \cite{PAVER}, which expresses 
the good
SU(3) symmetry of the couplings as intuitively expected.
Therefore, we can deduce:
\beq
\Gamma (S_3\rar K^+K^-)\simeq (73\pm 27)~\mbox{MeV} ~~~\mbox{and}~~~
\Gamma (S_3\rar \eta\eta)\simeq (15\pm 6)~\mbox{MeV}.
\eeq
A comparison of this result with the $f_0(1.5)$ width into $K\bar K$ of 5 MeV, 
and the
strong coupling of the $f_0(1.5)$ to $\pi\pi$ leads us
to the conclusion that this state cannot be a pure $\bar ss$ state.\\
We estimate the $S_3\gamma\gamma$
width using its relation 
with the one of $S_2$ and the corresponding quark charge:
\beq
\Gamma(S_3\rar\gamma\gamma)\approx 
\frac{2}{25}\ga\frac{M_{S_3}}{M_{S_2}}\dr^3\Gamma(S_2\rar\gamma\gamma)\approx
(0.40\pm 0.04)~\mbox{keV}.
\eeq
Finally, analogous to the case of $S'_2$, we can also have for 
$M_{S'_3}\simeq 1.7$ GeV:
\beq
\Gamma(S'_3\rar\gamma\gamma)\approx (1.1\pm 0.5)~\mbox{keV},~~~~~~~~~
\Gamma(S'_3\rar K^+K^-)\approx (112\pm 50)~\mbox{MeV}.
\eeq

\section{``Mixing-ology'' for the decay widths of scalar mesons}
We have discussed in the previous sections the masses and widths of the
unmixed gluonia and quarkonia states. The small value of the mass mixing
angle computed in \cite{PAK} from the off-diagonal
two-point correlator (see Fig. 11), which is proportional to the
light quark mass, 
allows us to neglect the off-diagonal term in the
mass matrix, and to identify the physical meson masses with
the ones of the unmixed states.

\begin{figure}[H]
\begin{center}
\includegraphics[width=10cm]{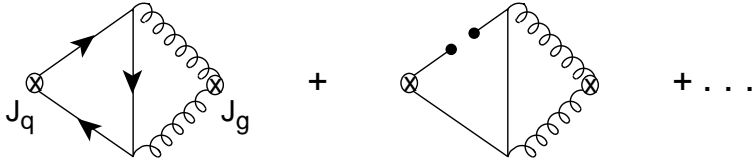}
\caption{Off-diagonal two-point correlator controlling
the quarkonium-gluonium mass mixing angle.}
\end{center}
\end{figure}

\nin
In the following, we shall be concerned
with the mixing angle for the couplings, which, in
the same approach, is controlled by the off-diagonal
non-perturbative
three-point function (see Fig. 12) which can (a priori) give a
large mixing angle.

\begin{figure}[H]
\begin{center}
\includegraphics[width=6cm]{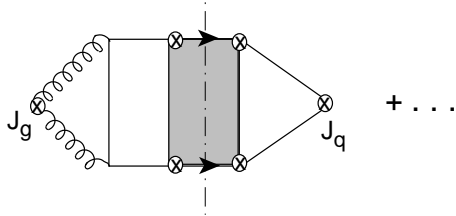}
\caption{Off-diagonal non-perturbative three-point function
controlling the coupling mixing angle in the quarkonium-gluonium
decays into pion pairs. }
\end{center}
\end{figure}

\nin
\subsection{Mixing below 1 GeV and the nature of the $\sigma$ and $f_0(0.98)$}
This part will be an update of the scheme proposed by BN \cite{BRAMON2}. 
We consider that the physically observed $f_0$ and $\sigma$ states
result from the two-component mixing of the $\sigma_B$ and $S_2\equiv
\frac{1}{\sqrt{2}}(\bar uu +\bar dd)$ unmixed bare states:
\bea
|f_0\ra &\equiv& -\sin\theta_S|\sigma_B\ra+\cos\theta_S|S_2\ra\nnb\\
|\sigma\ra &\equiv&~~\cos\theta_S|\sigma_B\ra+\sin\theta_S|S_2\ra .
\eea
We also use the previous prediction for
$\Gamma({\sigma_B}\rar\gamma\gamma)\simeq (0.2\sim 0.3)~\mbox{keV},$
and the experimental width $\Gamma(f_0\rar\gamma\gamma)\approx 0.3$ keV.
Therefore, we can fix the mixing angle to be:
\beq
\theta_S\approx (40\sim 45)^\circ,
\eeq
which indicates that the $f_0$ and $\sigma$ have a large amount of gluons in
their wave function. This situation is quite similar to the case of 
the $\eta'$
in the pseudoscalar channel (mass given by its gluon component, but 
strong coupling to quarkonia).  
Using the previous value of $\theta_S$, the predicted value of $g_{S_2K^+K^-}$,
the approximate relation $g_{S_2K^+K^-}\simeq \frac{1}{2}g_{S_2\pi^+\pi^-}$,
and the almost universal coupling of the $\sigma_B$ to pairs of Goldstone 
bosons, one can deduce:
\bea
 g_{f_0\pi^+\pi^-}&\simeq& (0.1\sim 2.6)~\mbox{GeV},~~~~~~g_{f_0K^+K^-}\simeq 
-(1.3\sim 4.1)~\mbox{GeV},\nnb\\
 g_{\sigma\pi^+\pi^-}&\simeq& ~g_{\sigma K^+K^-}\simeq
(4\sim 5)~\mbox{GeV},
\eea
which provides a simple explanation of the exceptional property of the $f_0$ 
(strong coupling to $\bar KK$ as observed in $\pi\pi$
and $\bar KK$ data \cite{MORGAN}), 
without appealing to the more exotic four-quark states and $\bar KK$
molecules \cite{BARNES}
\footnote{A QSSR analysis of the $a_0(0.98)$ within a four-quark
scheme leads to too low a value of its $\gamma\gamma$ width as
compared with the data \cite{FOUR}.}.  
Using the previous predictions for the couplings,
and for $\theta_S$, we obtain for $M_{\sigma_B}=(0.75\sim 1.0)$ GeV:
\beq
\Gamma(f_0(0.98)\rar\pi^+\pi^-)\approx (0.2\sim 134)~\mbox{MeV}~~~~~~~~~~~~~~~~~~~~~
\Gamma(\sigma\rar\pi^+\pi^-)\approx (300\sim 700)~\mbox{MeV},
\eeq
By recapitulating, our scheme suggests that around 1 GeV, there are two mesons
that have 1/2 gluon and 1/2 quark in  their wave functions
resulting from a maximal
destructive mixing between a quarkonium ($S_2$) and gluonium
$(\sigma_B)$ states:\\
$\b$ The $(a_0(0.98))$ is narrow, with a width $\leq$ 134 MeV,
and couples strongly to $\bar KK$, with 
the strength \\$g_{f_0K^+K^-}/g_{f_0\pi^+\pi^-}\simeq 2$. This property
has been seen in $\pi\pi$ and $\gamma\gamma$ scatterings \cite{MORGAN}
and in $\bar pp$ \cite{GASTALDI} experiments, and also suggests its
production from the $\phi$ radiative decay.
\\
$\b$
The $\sigma$, with a mass around $0.75\sim 1$ GeV,
is large, with a width of about $(300\sim 700)$ MeV.
\subsection{Mixing above 1 GeV 
and the nature of the $f_0(1.37)$, $f_0(1.5)$ and $f_J(1.7)$}
\subsubsection{The data}
Let us recall  the experimental facts
\footnote{In the following we shall take the largest range of values
deduced from the ones (sometimes controversial) given by PDG \cite{PDG}.}.
 Coupled channels analysis
demonstrate that in addition to the broad $\sigma(1.0)$ below
1.2 GeV, one needs \footnote{The question whether there is one or
two $f_0(1.37)$ (one coupled strongly to $2\pi$ and the
other to $4\pi$) is not yet settled.}
the $f_0(1.37)$, with the following properties \cite{PDG,GASPERO}:
\beq
\Gamma(f_0(1.37)\rar \mbox{all})\simeq (150\sim 500)~\mbox{MeV}
~~~~~~~~\Gamma(f_0(1.37)\rar\gamma\gamma)\simeq
(5.4\pm 2.3)~\mbox{keV},
\eeq
(where the number for the $\gamma\gamma$ width is less reliable); however,
from the quoted branching ratios and extracting the $4\pi$  S-wave,
we deduce \cite{PDG}:
\beq
\Gamma(f_0(1.37)\rar\pi\pi)\le (15\sim 100)~\mbox{MeV},~~~~~~~~~~~~~~~~~~~~~~~~~
\Gamma(f_0(1.37)\rar (4\pi)_S)\simeq (46\sim 316)~\mbox{MeV}.
\eeq
For the $f_0(1.5)$, one has \cite{BARREL,LANDUA}:
\beq
\Gamma(f_0(1.50)\rar~\mbox{all})\simeq (100\sim 150)~\mbox{MeV},~~~~~~~~~~~~~~
~~~~~~\frac{\Gamma(f_0(1.50)\rar4\pi)}{\Gamma(f_0(1.50)\rar\pi\pi)}
\simeq (3.4\pm 0.8),
\eeq
while the partial widths of the $f_0(1.5)$ divided by the phase-space factor  
satisfy the ratios:
\beq
\pi\pi~:~\eta\eta~:~\eta\eta'~:~\bar KK\simeq 1~:~(0.27\pm 0.11)~:~
(0.19\pm 0.08)~:~(0.24\pm 0.09).
\eeq
By assuming that the branching ratios other than the previous ones
are negligible, we can deduce the experimental data:
\bea
\Gamma(f_0(1.50)\rar\pi\pi)&\simeq& (20\sim 31)~\mbox{MeV}~~~~~~~~~
\Gamma(f_0(1.50)\rar \bar KK)\simeq (3.6\sim 5.6)~\mbox{MeV}\nnb\\
\Gamma(f_0(1.50)\rar\eta\eta)&\simeq& (2.6\sim 3.3 )~\mbox{MeV}~~~~~~~~~~~~
\Gamma(f_0(1.50)\rar 4\pi^0)\simeq (68\sim 105)~\mbox{MeV},
\eea
and:
\beq
\Gamma(f_0(1.50)\rar\eta\eta')\simeq 1.3~\mbox{MeV}.
\eeq
In the region between 1 and 1.5 GeV, we shall be concerned with the 
four unmixed states: the radial excitations $S'_2(1.3)$ of the $S_2$ 
and $\sigma'_B(1.37)$
of the $\sigma_B$, the $S_3(1.47)$ and the $G(1.5)$. \\ 
\subsubsection{Nature of the $f_0(1.37)$}
By inspecting the
experimental data in the region around $1.37$ GeV, 
one can expect that the $f_0(1.37)$ 
which has a large $\gamma\gamma$ width, 
is a good candidate for being essentially composed by the quarkonium
$S'_2$. However, its large decay width into $4\pi$ indicates its large
gluon component from the $\sigma'_B$. Due to the almost degenerate
values of the two previous unmixed states, we expect that they will
mix in a maximal way, as in the case of their corresponding ground
states $S_2-\sigma_B$ of mass around 1 GeV. Therefore, we expect
that it is difficult to disentangle the effects of these
two resonances where:\\
$\b$ The $S'_2$ quarkonium state is (relatively) narrower
with a total width around 400 MeV. It decays mainly 
into $\pi\pi$ (its decays into $\bar KK$ and $\eta\eta$
can be obtained by taking proper Clebsch-Gordan factor
and assuming $SU(3)$ symmetry of the couplings.),
and has a large $\gamma\gamma$ width of about 4 keV.\\
$\b$ The $\sigma'$ is very broad, with a width as large as 1 GeV, 
has an universal coupling
to $\pi\pi$, $\bar KK$ and $\eta\eta$,
 and can decay into $4\pi$ via $\sigma\sigma$,
with a width of $(46\sim 316)$ MeV. However, its decay into $\gamma\gamma$
is relatively small.\\
$\b$ The experimental candidate $f_0(1.37)$ has amusingly the combined properties
of the $S'_2$ and $\sigma'$.\\
\subsubsection{A 3x3 mixing scheme and nature of the $f_0(1.5)$}
In the
following estimate of
the hadronic widths of the remaining other mesons, we expect that, due to
its large width, the
$\sigma'$ can have a stronger mixing with the $S_3(1.47)$ and the $G(1.5)$,
than the $S'_2$, such that, to a first approximation, 
we shall only consider the mixing of the three
states $\sigma'(1.37)$, $S_3(1.47)$ and the $G(1.5)$
through the unitary $3\times 3$ CKM-like
mixing matrix of the couplings \footnote{In the following,
we shall neglect the $CP$-violating phase. It can also be noticed that
we cannot fix the masses of the physical states from this mixing of
the couplings. However, as shown earlier from the small mass-mixing
angle, we expect
that their masses are around those of the unmixed states.}:

\nin
\\
\beq\left(
\begin{array}{c}
f_0(1.37)\\
f_0(1.50)\\
f_0(1.60)\\
\end{array}
\right)
=
\left(
\begin{array}{ccc}
c_{12}c_{13}&s_{12}c_{13}&s_{13}\\
-s_{12}c_{23}-c_{12}s_{23}s_{13}&c_{12}c_{23}-s_{12}s_{23}s_{13}&s_{23}c_{13}\\
s_{12}s_{23}-c_{12}c_{23}s_{13}&-c_{12}s_{23}-s_{12}c_{23}s_{13}&c_{23}c_{13}\\
\end{array}
\right)
\left(
\begin{array}{c}
\sigma'(1.37)\\
S_3(1.47)\\
G(1.5)\\
\end{array}
\right)
\eeq
\nin
\\
where:
\beq
c_{ij}=\cos{\theta_{ij}}~~~~\mbox{and}~~~~s_{ij}=\sin{\theta_{ij}}.
\eeq
We shall use in our numerical analysis:
\bea
\Gamma(\sigma'\rar\pi\pi)&\approx& (1\sim 2)~\mbox{GeV}~~~~~~~~~
\Gamma(\sigma'\rar 4\pi)\approx (40~\sim 316)~\mbox{MeV  (data)},\nnb\\
\Gamma(\sigma'\rar\bar KK)&\approx& (0.7\sim 1.5)~\mbox{GeV}~~~~~~~~~
\Gamma(\sigma'\rar\eta\eta)\approx (0.2\sim 0.4)~\mbox{GeV},
\eea
and we shall discuss separately the cases of large (upper values) and small
(lower values) of each partial widths.\\

\nin
$\b$ {\bf The case of small widths}\\
We estimate each entry as follows:\\
For the first line of the matrix,
 we shall fix $c_{12}$ from
the small width of $f_0(1.37)$ into $\bar KK$, which gives:
\beq
c_{12}\approx 0.31~~~~~~~~ s_{12}\approx -0.95.
\eeq
Then , we fix $\theta_{13}$, by requiring the best
prediction:
\beq
\Gamma(f^B_0(1.37)\rar 4\pi)\approx 150~\mbox{MeV}
\eeq
compared with the previous data. This implies the two solutions (a) and (b):
\beq
c_{13}\approx~~~\mbox{(a)}~~~-0.29 ~~~~~~~\mbox{and}~~~\mbox{(b)}~~~~~~0.74.
\eeq
For the second line of the matrix, we use the observed width
$\Gamma(f_0(1.5)\rar \pi\pi)$, in order to get:
\beq
c_{23}\approx~~~ \mbox{(a)}~~~0.45 ~~~~~~~\mbox{and}~~~\mbox{(b)}~~~~~~0.37.
\eeq
for the corresponding two values of $c_{13}$. Finally, the observed width 
$\Gamma(f_0(1.5)\rar 4\pi)$ favours the alone case (b).\\
\nin
$\b$ {\bf The case of large widths}\\
We repeat the previous analysis taking the upper values
of the partial widths.\\
\nin
$\b$ {\bf Predictions}\\
Our final mixing matrix is given by the largest range spanned by each mixing
angles. Therefore, the mixing matrix reads:
\\
\beq\left(
\begin{array}{c}
f_0(1.37)\\
f_0(1.50)\\
f_0(1.60)\\
\end{array}
\right)
\approx
\left(
\begin{array}{ccc}
0.01\sim 0.22&-(0.44\sim 0.7)&0.89\sim 0.67\\
0.11\sim 0.16&0.89\sim 0.71&0.43\sim 0.69\\
-(0.99\sim 0.96)&-(0.47\sim 0.52)&0.14\sim 0.27\\
\end{array}
\right)
\left(
\begin{array}{c}
\sigma'(1.37)\\
S_3(1.47)\\
G(1.5)\\
\end{array}
\right),
\eeq

\nin
where the first (resp. second) numbers correspond to the case of large (resp. small)
widths. From the previous schemes, we deduce the predictions
{\footnote{Recall that we have used as inputs:
$\Gamma(f_0(1.37)\rar\bar KK)\approx 0,~
\Gamma(f_0(1.5)\rar\pi\pi)\approx 25~\mbox{MeV}$,
while our best prediction for $\Gamma(f_0(1.5)\rar (4\pi)_S)$
is about 150
MeV. The present data also favour negative values of the $f_0\eta\eta$,
$f_0\eta'\eta$ and $f_0KK$ couplings.}:
\beq
\Gamma(f_0(1.37)\rar\pi\pi)\approx (22\sim 48)~\mbox{MeV}~~~~~~~~~
\Gamma(f_0(1.37)\rar\eta\eta)\leq 1.~\mbox{MeV}~~~~~~~~~
\Gamma(f_0(1.37)\rar \eta\eta')\leq 2.5 ~\mbox{MeV}
\eeq
and
\bea
\Gamma(f_0(1.5)\rar\bar KK)\approx (3\sim 12)~\mbox{MeV}~~~~~~
\Gamma(f_0(1.5)\rar\eta\eta)\approx (1\sim 2)~\mbox{MeV}~~~~~~
\Gamma(f_0(1.5)\rar\eta\eta')\leq 1.~\mbox{MeV}~.
\eea
Despite the crude approximation used and the inaccuracy of the
predictions, these results are
in good agreement with the data, especially from the
Crystal Barrel collaboration. These results suggest that the 
observed $f_0(1.37)$ and $f_0(1.5)$ come from a maximal mixing
between the gluonia ($\sigma'$ and $G$) and the quarkonium $S_3$
states. The mixing of the $S_3$ and $G$
with the quarkonium $S'_2$, which we have neglected compared with
the $\sigma'$, can restore the small discrepancy with the data.
One should notice, as already mentioned, that the state seen 
by GAMS is more likely to be similar to the unmixed
gluonium state $G$ (dominance of the
$4\pi$ and $\eta\eta'$ decays,
as emphasized earlier in NV), which can be due to some specific
features of the central production (double pomeron exchange mechanism
which favours the gluonia production) for the GAMS experiment. This feature
is not shared by the data from the Crystal Barrel and Obelix collaborations,
which correspond to the annihilation of antiprotons at rest in a hydrogen
target, and therefore can also favour the production of quarkonia states
\footnote{We plan to come back in more details to this point in a future work.}.
\subsubsection{Nature of the $f_J(1.71)$}
For the $f_0(1.6)$, we obtain:
\bea
\Gamma(f_0(1.6)\rar \bar KK)&\approx& (0.5\sim 1.6)~\mbox{GeV}~~~~~~~~~
\Gamma(f_0(1.6)\rar\pi\pi)\approx (0.9\sim 2.)~\mbox{GeV}
\nnb\\
\Gamma(f_0(1.6)\rar\eta\eta)&\approx& (0.04\sim 0.6)~\mbox{GeV}~~~~~~~~
\Gamma(f_0(1.6)\rar \eta\eta')\approx (0.03\sim 0.07)~\mbox{GeV},
\eea
and
\beq
\Gamma(f_0(1.6)\rar (4\pi)_S)\approx (0.02\sim 0.2)~\mbox{GeV},
\eeq
which suggest that the
$f_0(1.6)$ is very broad and can again be confused with the continuum.
Therefore, the $f_J (1.7)$ observed to
decay into $\bar KK$ with a width of the order $(100\sim 180)$ MeV,
can be essentially composed by the radial excitation 
$S'_3(1.7\sim 2.4)$ GeV of the $S_3(\bar ss)$, as they have
about the same width into $\bar KK$ (see section 6.4). This can
also explain
the smallness of the $f_J(1.7)$ width into $\pi\pi$ and $4\pi$.
Our predictions of the $f_J(1.71)$ width can agree with
the result of the Obelix collaboration \cite{GASTALDI}, while
its small decay width into $4\pi$ is in agreement with the
best fit of the Crystal Barrel collaboration 
(see Abele et al. in \cite{BARREL}), which is consistent with
the fact that the $f_0(1.37)$ likes to decay into 4$\pi$.
However, the $f_0(1.6)$ and the $f_J(1.71)$ can presumably
interfere destructively for giving the dip around 
$1.5\sim 1.6$ GeV seen in 
the $\bar KK$ mass distribution
from the Crystal Barrel and $\bar pp$
annihilations at rest \cite{ABELE,GASTALDI}. 
\subsubsection{Comparison with other scenarios}
One can also compare our
results with some other
mixing scenarios \cite{CLOSE2, LAHIRI}. Though 
the relative amount of glue for the $f_0(1.37)$ and $f_0(1.5)$
is about the same here and in \cite{CLOSE2} , one should notice
that, in our case, the $\pi\pi$ partial width of these mesons come mainly
from the $\sigma'$, a glue state coupled strongly to the
quark degrees of freedom, like the $\eta'$ of the $U(1)_A$ anomaly,
while in \cite{CLOSE2}, the $S_2$ which has a mass higher than
the one obtained here plays an essential role in the mixing.
Moreover, the $f_J(1.71)$ differs significantly in the two approaches, as
here, the $f_J(1.71)$ is mainly the $\bar ss$ state $S'_3$, 
while in \cite{CLOSE2},
it has a significant gluon component. In the present approach, the eventual
presence of a large gluon component into the $f_J(1.71)$ wave function
can only come from the mixing
with the broad $f_0(1.6)$ and with
the radial excitation of the gluonium $G$(1.5), which mass
is expected to be around 2 GeV (value of $\sqrt{t_c}$ in the QSSR
analysis). However, the apparent absence
of the $f_J(1.71)$ decay into 4$\pi$ from Crystal Barrel data may not
favour such a scenario.
\subsubsection{Summary}
Within the present approach and the present data, 
we expect above 1 GeV that:\\
$\b$ the $f_0(1.37)$ is a superposition of two states: the radial 
excitation $S'_2$ of the quarkonium $S_2\equiv (\bar uu+\bar dd)$ and
an $f_0(1.37)$ resulting from a maximal mixing between
the gluonium $G$  and the $S_3(\bar ss)$
with the radial excitation $\sigma'_B$ of the broad low mass $\sigma_B$.
Its large $\gamma\gamma$ width comes from the $S'_2$, while its
affinity to decaying into $4\pi$ comes essentially from the $\sigma'$.\\
$\b$ the $f_0(1.5)$ with the properties observed by the Crystal
Barrel collaboration results from a maximal mixing between
the gluonium $G$  and the $S_3(\bar ss)$
with the radial excitation $\sigma'$ of the broad low-mass $\sigma$.
This large gluon component explains its affinity
to decaying into $2\pi$ (from $\sigma'(1.37)$) and $4\pi$ 
(from $\sigma'$ and $G$)\\
$\b$ the $f_J(1.71)$ can be identified with the
radial excitation  $S'_3$ of the ground state $S_3(1.47)$,
as they have about the same width into $\bar KK$. This can also explain 
the absence of its decay into $\pi\pi$ and 4$\pi$. The eventual
observation of these  decays can measure its expected tiny
mixing with the wide (and presumably unobservable)  $f_0(1.6)$ and with
the radial excitation of the $G(1.5)$. The presence of the dip 
around $(1.5\sim 1.6)$ GeV seen in the $\bar KK$ invariant 
mass distribution may already be a signal of a such mixing. 
However, we need a clear
spin-parity assignement of the $f_J(1.71)$ before a definite conclusion
on its nature can be drawn.
\section{The tensor gluonium}
\subsection{Mass and decay constant}
In this section, we shall study the properties of the $2^{++}$ gluonium.
We shall be concerned with the two-point correlator:
\bea
\psi^T_{\mu\nu\rho\sigma}&\equiv& i\int d^4x~ e^{iqx}\la 0|{\cal T}
\theta^g_{\mu\nu}(x)
\theta^g_{\rho\sigma}(0)^\dagger|0\ra\nnb\\
&=&\frac{1}{2}
\ga \eta_{\mu\rho}\eta_{\nu\sigma}+\eta_{\mu\sigma}\eta_{\nu\rho}
-\frac{2}{3}\eta_{\mu\nu}\eta_{\rho\sigma}\dr \psi_T(q^2),
\eea
where:
\beq
\eta_{\mu\nu}\equiv g_{\mu\nu}-\frac{q_\mu q_\nu}{q^2}.
\eeq
To leading order in $\alpha_s$ and including the non-perturbative condensates,
the QCD expression of the correlator reads \cite{NSVZ}:
\beq
\psi_T(q^2\equiv -Q^2)= -\frac{1}{20\pi^2}(Q^4)\log{\frac{Q^2}{\nu^2}}
+\frac{5}{12}\frac{g^2}{Q^4}\la 2O_1-O_2\ra~,
\eeq
where:
\beq
O_1=\ga f_{abc}G_{\mu\alpha}G_{\nu\alpha}\dr^2~~~~~~~~\mbox{and}~~~~~~~~
O_2=\ga f_{abc}G_{\mu\nu}G_{\alpha\beta}\dr^2.
\eeq
Using the vacuum saturation hypothesis, one can write:
\beq
\la 2O_1-O_2\ra\simeq -\frac{3}{16}\la G^2\ra^2.
\eeq
 In order to get the gluonium mass, we work with the ratio of moments:
\beq
{\cal R}_{10}\equiv \frac{{\cal L}_T^{(1)}}{{\cal L}_T^{(0)}},
\eeq
as in SN \cite{SNG}. Its QCD expression reads:
\beq
{\cal R}_{10}\simeq {3\tau^{-1}}\aga 1+\frac{25}{24}\pi^3
\frac{\tau^4}{\alpha_s}\gg^2\adr.
\eeq

\begin{figure}[H]
\begin{center}
\includegraphics[width=6cm]{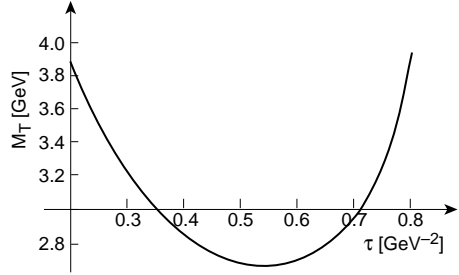}
\caption{$\tau$ behaviour for the upper bound of the tensor 
gluonium mass $M_T$.}
\end{center}
\end{figure}

\nin
Using the positivity of the spectral function, the minimum of the ratio
of the moments (Fig. 13) leads to the upper bound:
\beq
M_T\le (2.7\pm 0.4)~\mbox{GeV},
\eeq
while a resonance + QCD continuum parametrization of the spectral function
gives the value (Fig. 14):
\beq
M_T\simeq (2.0\pm 0.05\pm 0.05\pm 0.05)~\mbox{GeV}\simeq (2.0\pm 0.1)~
\mbox{GeV}.
\eeq

\begin{figure}[H]
\begin{center}
\includegraphics[width=10cm]{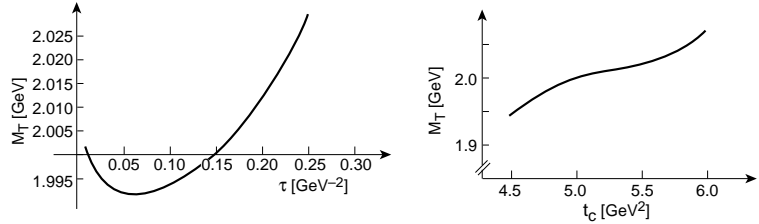}
\caption{a) $\tau$ and b) $t_c$ behaviours of the $2^{++}$
tensor gluonium mass.}
\end{center}
\end{figure}

\nin
The mass obtained here is larger than the one in SN, which is mainly due to
the increase of the gluon condensate value.
The errors come respectively from the gluon condensate, the factorization
assumption and the continuum threshold $t_c$. The decay constant
can be extracted from ${\cal L}_T^{(0)}$, and reads (Fig. 15):
\beq
f_T\simeq (80\pm 8\pm 4\pm 4\pm 10)~\mbox{MeV}\simeq (80\pm 14)~\mbox{MeV},
\eeq
where the first error comes from $M_T$ and the remaining ones are of the
same origin as before.

\begin{figure}[H]
\begin{center}
\includegraphics[width=6cm]{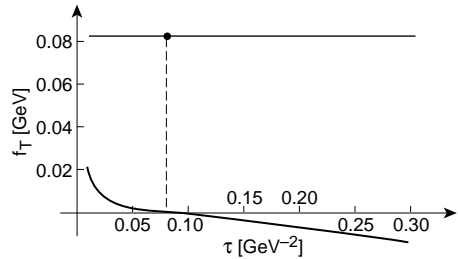}
\caption{$\tau$ behaviour of the $2^{++}$ tensor gluonium
decay constant $f_T$. }
\end{center}
\end{figure}

\nin
The value of $t_c$ at which 
these results are optimal is:
\beq
\sqrt{t_c}\approx M_{T'}\simeq (2.2\sim 2.3)~\mbox{GeV},
\eeq
which corresponds roughly to the mass of the radial excitation.
Our result is slightly higher than the one in SN and
in \cite{DOMI}, because of the increase of the
value of the gluon condensate. We do not expect that the perturbative
radiative corrections
will affect noticeably our prediction for the 
mass \footnote{An investigation of these effects
with K. Chetyrkin and A. Pivovarov is under way. A preliminary result
indeed indicates that the effects are small.} , as this effect tends
to cancel out in the ratio of moments.
\subsection{Tensor gluonium decay widths}
Using the result in \cite{BRAMON}, one can constrain the ratio:
\beq
r\equiv \frac{g_{T\pi\pi}}{g_{f_2\pi\pi}}\simeq -(0.68\pm 0.23),
\eeq
using the data, where:
\beq
\Gamma({T\rar\pi^+\pi-})=\frac{g^2_{T\pi\pi}}{30\pi M^2_T}|k_\pi|^5,
\eeq
$k_\pi$ being the pion momentum.
 Assuming an universal
coupling of the $T$ to the pairs of Goldstone bosons,
this leads to the width:
\beq
\Gamma(T\rar\pi\pi+KK+\eta\eta)\leq (119\pm 36)~\mbox{MeV},
\eeq
which indicates that the $2^{++}$ cannot be wide, contrary to some
claims in the literature. Another indication of the smallness of
the $2^{++}$ width can be obtained from the low-energy theorem
analogue of the one for the scalar gluonium used in section 5
\cite{ALIEV}. Using a dipersion relation, one can write:
\beq
\frac{1}{4}\sum_{f_2,T}\sqrt{2}f_Tg_{T\pi\pi}\approx 1.
\eeq
As in \cite{BRANA}, we assume that the 
vertex is saturated by the $f_2(1.27)$ and by the gluonium $T$. Using
$f_{f_2} \approx 0.19$ GeV \cite{BAGAN3}, the previous value of $f_T$, and
the experimental value $g_{f_2\pi\pi} \simeq 16$ GeV$^{-1}$, one can notice
that the sum rule is already saturated by the $f_2$-meson, and leads to:
\beq
g_{T\pi\pi} \approx 2.6~\mbox{GeV}^{-1}\lrar \Gamma({T\rar\pi\pi})\approx
10~\mbox{MeV},
\eeq
while taking a typical 10$\%$ error in the estimate of $f_{f_2}$, 
one can deduce:
\beq
 \Gamma({T\rar\pi\pi})\leq 70~\mbox{MeV}.
\eeq
\\
The $\gamma\gamma$ width can be obtained by relating it to the one of the
$0^{++}$ gluonium within a non-relativistic quark model
approach \cite{BRAMON}:
\beq
\Gamma(T\rar\gamma\gamma)\simeq \frac{4}{15}\ga\frac{M_T}{M_{0^+}}\dr^3
\Gamma(0^{++}\rar\gamma\gamma)\simeq 0.06~\mbox{keV},
\eeq
which shows again a small value typical of a gluonium state. 
\subsection{Meson-gluonium mass mixing and the
nature of the $f_J(1.71)$ and $\zeta(2.2)$}
In order to evaluate the gluonium-quarkonium mass mixing angle, we
shall work, as in \cite{PAK, BRAMON}, with the off-diagonal
two-point correlator:
\bea
\psi^{gq}_{\mu\nu\rho\sigma}&\equiv& i\int d^4x~ e^{iqx}\la 0|{\cal T}
\theta^g_{\mu\nu}(x)
\theta^q_{\rho\sigma}(0)^\dagger|0\ra\nnb\\
&=&\frac{1}{2}
\ga \eta_{\mu\rho}\eta_{\nu\sigma}+\eta_{\mu\sigma}\eta_{\nu\rho}
-\frac{2}{3}\eta_{\mu\nu}\eta_{\rho\sigma}\dr \psi_{gq}(q^2),
\eea
where
\beq
\theta^q_{\mu\nu}(x)=i\bar q(x)(\gamma_\mu \overline{D}_\nu+\gamma_\nu 
\overline{D}_\mu)q(x).
\eeq
Here, $\overline{D}_\mu\equiv \overrightarrow{D}_\mu-
\overleftarrow{D}_\mu$ is the covariant derivative,
and the other quantities have already been defined earlier.
Taking into
account the mixing of the currents, one obtains \cite{BRAMON}:
\beq
\psi_{gq}(q^2\equiv -Q^2)\simeq \frac{q^4}{15\pi^2}\asr\ga 
\log^2{\frac{Q^2}{\nu^2}}-
\frac{91}{15}\log{\frac{Q^2}{\nu^2}}\dr -\frac{7}{36\pi}
\log{\frac{Q^2}{\nu^2}}\gg .
\eeq 
The resonance contribution to 
the spectral function is introduced using a two-component mixing formalism:
\beq
\frac{1}{\pi}\mbox{Im}\psi_{gq}(t)\simeq \sin{2\theta_T}M^2_Tf_TM^2_Ff_F
\delta
(t-M^2)~+~
\mbox{``QCD continuum"},
\eeq
where $\theta_T$ is the mixing angle; $F$ is a generic notation for the 
quarkonia
$f_2(1.27)$ and $f'_2(1.53)$; $M$ is the average of the $G$ and $F$ 
mass squared;
$f_{T,F}$ is the decay constant, where
$f_{f_2}\simeq (0.11\sim 0.14)M_{f_2}$ \cite{BRANA}. Noting that within our 
approximation,
the Laplace transform sum rule does not present a $\tau$ stability, 
we work with
the FESR:
\beq
\int_0^{t_c}dt~\frac{1}{\pi}\mbox{Im}\psi_{gq}(t)\simeq
-\frac{101}{675\pi^2}\as t^3_c\ga 1+\frac{525\pi^2}{404t^2_c}\la G^2\ra\dr ,
\eeq
where $t_c\simeq\frac{1}{2}(t^q_c+t^g_c)\simeq 2.2$ GeV$^2$ is the average of 
the continuum
threshold for the quark and gluonia channels. Then, one obtains 
the approximate estimate \cite{BRAMON}
\footnote{One should notice that, to this approximation, we do not have $t_c$
stability, such that our result should only be taken as a crude indication, but not
as a precise estimate.}:
\beq
\theta_T\simeq -10^\circ,
\eeq
which is a small mixing angle, and which indicates that:\\
$\b$ the interpretation of the $f_J(1.71)$ as the
lowest 2$^{++}$ ground state gluonium, is not favoured by 
our present result.\\
$\b$ the $\zeta(2.2)$ observed by the BES collaboration
\cite{BES} is a good gluonium candidate (mass and width), 
but it can be more probably the first radial excitation of
the lowest-mass gluonium ($t_c \approx 2.2$ GeV). Our result, for the mass,
suggests that one needs further data around 2 GeV for finding 
the lowest mass $2^{++}$ gluonium.
Its (non-)finding will be a test for the accuracy of
our approach. Our prediction for the mass and for $t_c$, which 
suggests a rich population of $2^{++}$
states in this region, should stimulate a further independent experimental test
of the $g_T$ candidates seen earlier by the BNL group in the OZI suppressed
reaction $\pi^- p\rar \phi\phi n$ \cite{LINDEN}, and a further understanding of
their non-observation in $J/\psi$ radiative decays \cite{HITLIN}.
\section{Pseudoscalar gluonia}
\subsection{Mass and decay constant}
The pseudoscalar gluonium sum rules have been discussed many times in
the literature, in connection with the gluonium mass in SN \cite{SNG},
$\eta'$ meson, the spin
of the proton and the slope of the topological $U(1)$ charge in
\cite{SHORE,SNG2}. Here, we update the analysis of the gluonium mass
in SN, taking into account the
new, large perturbative radiative corrections \cite{KATAEV}, 
and the radiative correction for the gluon condensate \cite{STEELE}. 
In so doing we work with the Laplace transform:
\bea
{\cal L}_P^{(0)}&\equiv&\int_{t_\leq}^{\infty} {dt}~\mbox{exp}(-t\tau)
~\frac{1}{\pi}~\mbox{Im} \psi_P(t)\nnb\\
&=&\ga\frac{1}{8}\dr^2\as^2\frac{4}{\pi^2}\tau^{-3}
\Bigg{[}\ga 1+\delta_{pert}\as\dr\ga 1-\rho_2\dr\nnb\\
&&-\frac{11\pi}{4}\gg\tau^2\ga 1-\exp(-t_c\tau)\dr
-2\pi^2\tau^3g\la f_{abc}G^3_{abc}\ra\Bigg{]},
\eea
where:
\bea
\delta_{pert}
&\equiv& \frac{1}{4}\Bigg{[} 83+2\beta_1(1-2\gamma_E)-4\frac{\beta_1}
{\beta_2}
\log(-\log{\tau\Lambda^2})\Bigg{]}\nnb\\
\rho_2&\equiv& \ga 1+t_c\tau+\frac{(t_c\tau)^2}{2}\dr\exp({-t_c\tau})
\eea
and the ratio of moments
\beq
{\cal R}_{10}\equiv
-\frac{d}{d\tau}\log{{\cal L}_P^{(0)}}
\equiv \frac{{\cal L}_P^{(1)}}{{\cal L}_P^{(0)}}.
\eeq
Subtracting the $\eta'$ contribution\footnote{In order
to take into account the
change in the radiative correction coefficient, we have redone the
estimate of the $\eta'$ decay constant obtained in \cite{SHORE};
we obtain a slight change $f_{\eta'}\simeq 30$ MeV in agreement,
within the error, with the previous value of $(24\pm 3.5$) MeV. 
We have used in the
massless quark limit $\tilde{M}^2_{\eta'}=M^2_{\eta'}-\frac{2}{3} M^2_K
\simeq (0.87~\mbox{GeV})^2$.}, and 
using the positivity of the spectral function, the minimum of ${\cal R}_{10}$
gives the upper bound (Fig. 16):
\beq
M_P\le (2.34\pm 0.42)~\mbox{GeV},
\eeq
where the error comes mainly from the value of $\Lambda$, which one can 
understand as the minimum occurring at large values of $\tau$.

\begin{figure}[H]
\begin{center}
\includegraphics[width=6cm]{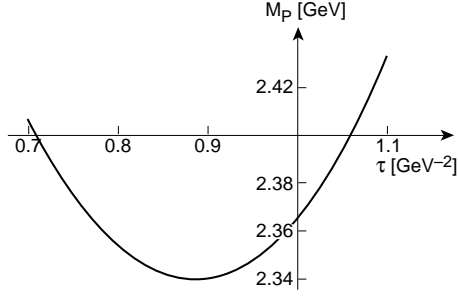}
\caption{$\tau$ behaviour of the
upper bound for the pseudoscalar gluonium mass $M_P$.}
\end{center}
\end{figure}

\nin
The estimate of the pseudoscalar
gluonia mass is given  in Fig. 17, where the stability
in $\tau$ is obtained at smaller $\tau$ values, which gives smaller
 errors. We obtain:
\beq
M_P\simeq (2.05\pm 0.05\pm 0.13\pm 0.13)~\mbox{GeV}
\simeq (2.05\pm 0.19)~\mbox{GeV},
\eeq
where the first error comes from $t_c$, the second and the third ones come
from $\Lambda$ and the truncation of the QCD series. As in the previous
cases, we have estimated the unknown $\alpha_s^2$ coefficient to be of the
 order of $\pm$400 (normalized to the lowest order term)
by assuming a geometric growth of the QCD series. 

\begin{figure}[H]
\begin{center}
\includegraphics[width=10cm]{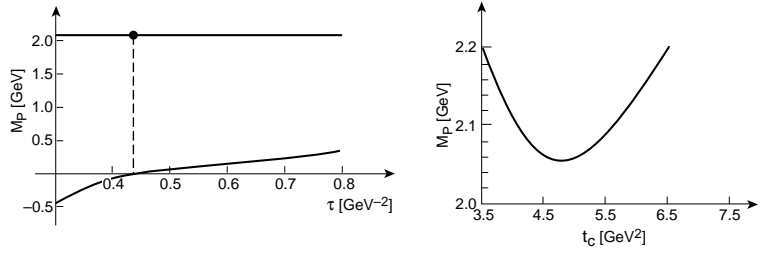}
\caption{$\tau$ and $t_c$ behaviours of the 
pseudoscalar gluonium mass.}\end{center}
\end{figure}

\nin
The corresponding value of $t_c$ is:
\beq
\sqrt{t_c}\approx M_{P'}\simeq (2.1\sim 2.3)~\mbox{GeV}.
\eeq
The decay constant has a good $\tau$ stability (Fig. 18) though
the $t_c$ stability is only reached at $t_c\simeq 7$ GeV$^2$ (Fig. 21). Using
the range of $t_c$ values from 5 to 7 GeV$^2$, we deduce:
\beq
f_P\simeq (8\sim 17)~\mbox{MeV}.
\eeq

\begin{figure}[H]
\begin{center}
\includegraphics[width=10cm]{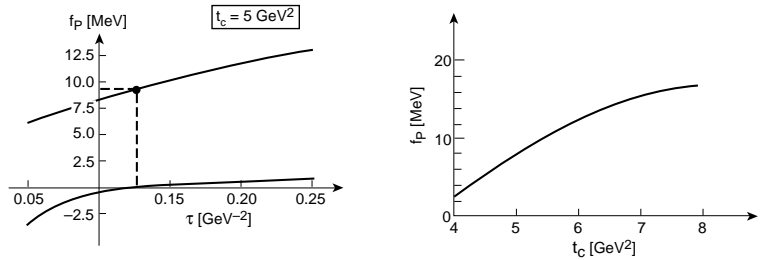}
\caption{ $\tau$ and $t_c$ behaviours of the decay constant
$f_P$ for $t_c=5.5$ GeV$^2$.}
\end{center}
\end{figure}

\nin
These results are slightly higher than the one in SN obtained from a numerical 
least-squares fit, because mainly of the effect of the radiative corrections.
\subsection{Testing the nature of the $E/\iota(1.44)$}
Old \cite{PDG} and new \cite{MONTA} experimental data indicate the presence 
of some extra states in the 
range of 1.40-1.45 GeV, which go beyond the usual nonet
classification. In order, to test the nature of such states (denoted hereafter
as the $E/\iota(1.44)$), let us now come back to the sum rule:
\beq
{\cal L}_P^{(-1)}\equiv\int_{t_\leq}^{\infty}\frac{dt}{t}~\mbox{exp}(-t\tau)
~\frac{1}{\pi}~\mbox{Im} \psi_P(t),
\eeq
which has been used in \cite{SHORE} for fixing the decay constant $f_{\eta'}$:
\beq
\la 0| Q(x)|\eta'\ra =\sqrt{2}f_{\eta'}M^2_{\eta'}.
\eeq
By defining in the same way the decay constant $f_\iota$, we introduce
into the sum rule the parameters of the $\eta'$ and $P$ gluonium and the
corresponding value of $t_c\simeq 6\sim 7$ GeV$^2$ at which $f_{\eta'}$
has been optimized. In this way, one finds that there is no room to
include the $\iota$ contribution, as:
\beq
f_\iota\approx 0. 
\eeq
One can weaken
the constraint by replacing the QCD continuum effect, i.e.
all higher-state effects, by the one of the 
$\iota$, which should lead to an overestimate of $f_\iota$. 
In this way, one can deduce the upper bound (Fig. 19):
\beq
f_{\iota} \le 16~\mbox{MeV}.
\eeq

\begin{figure}[H]
\begin{center}
\includegraphics[width=6cm]{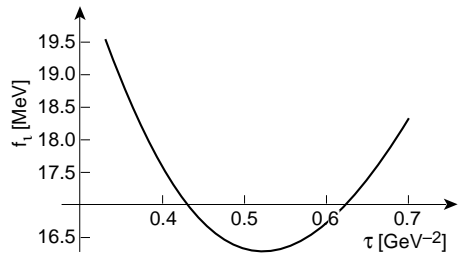}
\caption{$\tau$ behaviour of the decay constant
$f_\iota$ for $t_c=6.$ GeV$^2$ }
\end{center}
\end{figure}

\nin
One can compare the previous result with the one obtained from $J/\psi$
radiative decays, which gives \cite{NSVZ2}:
\beq
B_{\iota \eta'}
\equiv \frac{\Gamma(J/\psi \rar \gamma\iota)}{\Gamma(J/\psi \rar \gamma\eta')}
\simeq \Bigg{\vert}\frac{\la 0| Q(x)|\iota\ra}{\la 0| Q(x)|\eta'\ra}
\Bigg{\vert}^2
\ga\frac{k_\iota}{k_{\eta'}}\dr^3,
\eeq
where the matrix element is controlled by the decay constant of the 
corresponding particle. Using the
experimental branching ratio, where we take the one for $\iota\rar \bar KK\pi$,
we deduce:
\beq
f_{\iota} \simeq 0.23 f_{\eta'}\simeq 7~\mbox{MeV},
\eeq
in agreement with our findings. 
Our analysis indicates that the $E/\iota$ couples more weakly to 
the gluonic
current than the $\eta'$ ($f_{\eta'}\approx 30
~\mbox{MeV}$), and is thus
likely to be the radial excitation of the $\eta'$ \footnote{
The $E/\iota$ has been interpreted as a bound state of light gluinos
\cite{FARRAR}. However, we can consider that the results in \cite{CHET2}
do not favour this interpretation, as
the inclusion of the light gluino loops leads to an overestimate of 
the value of $\alpha_s(M_Z)$ from $\tau$-decays \cite{BNP}--\cite{PICH}
compared with the world average \cite{BETHKE}. This conclusion being confirmed
from a global fit
analysis of the $Z$ and $\tau$ hadronic decays \cite{FODOR}, and from the new
ALEPH lower limit of the gluino mass of about 6.3 Gev ($95\%$ CL)
from the running of $\alpha_s$
and four-jet variables \cite{DISSERTORI}.  
We also learn from U. Gastaldi that the $\bar pp$
data on $\bar KK\pi$
can be fitted by one resonance, though the quality of the fit is
obviously much better within a two-resonance fit.}.

\subsection{Meson-gluonium mixing and $P\rar\gamma\gamma,\rho\gamma$ decays}
Following \cite{PAK}, we obtain from the evaluation of the
off-diagonal two-point correlator, the quarkonium-gluonium mixing angle 
\cite{PAK,SNB}
\beq
\theta_P\simeq 12^\circ,
\eeq
from which one can deduce the decay widths of the pseudoscalar gluonium $G$:
\bea
\Gamma(P\rar\gamma\gamma)&\simeq& \tan^2{\theta_P}
\ga\frac{M_P}{M_{\eta'}}\dr^3\Gamma(\eta'\rar\gamma\gamma)
\approx (1.3\pm 0.1)~\mbox{keV}\nnb\\
\Gamma(P\rar\rho\gamma)&\simeq& \tan^2{\theta_P}
\ga\frac{k_P}{k_{\eta'}}\dr^3\Gamma(\eta'\rar\rho\gamma)\approx 
(0.3\pm 0.1)~\mbox{MeV},
\eea
where $k_i$ is the momentum of the particle $i$. 
We have used $\Gamma(\eta'\rar\gamma\gamma)\simeq 4.3~\mbox{keV}$ and
$\Gamma(\eta'\rar\rho\gamma)\simeq (72\sim 21)~\mbox{keV}$. Measurements of the $P$ 
widths
can test the amount of glue inside the $P$-meson.

\section{Conclusions}
$\b$ We have updated the QCD spectral sum rule (QSSR) analysis
for computing the masses and
decay constants of the scalar (section 4),
tensor (section 8) and pseudoscalar (section 9) gluonia, using the present
values of the QCD parameters and the recent progresses in the QCD evaluation
of the gluonic correlators.
The results of the analysis are summarized in Table 1:
\begin{table}[H]
\begin{center}
\begin{tabular}[h]{llllll}
\hline 
  & & & & &\\
$J^{PC}$&Name&\multicolumn{2}{l}{Mass [GeV]}&$f_G$ [MeV]&$\sqrt{t_c}$ [GeV]\\
   & & & & &\\
&&Estimate&Upper Bound&&\\
&&&&&\\
\hline 
 & & & & & \\
$0^{++}$&$G$&$1.5\pm 0.2$&$2.16\pm 0.22$&$390\pm 145$&$2.0\sim 2.1$\\
&$\sigma_B$&1.00 (input)&&1000&\\
&$\sigma'_B$&1.37 (input)&&600&\\
&$3G$&3.1&&62&\\
&&&&&\\
$2^{++}$&$T$&$2.0\pm 0.1$&$2.7\pm 0.4$&$80\pm 14$&$2.2$\\
&&&&&\\
$0^{-+}$&$P$&$2.05\pm 0.19$&$2.34\pm 0.42$&$8\sim 17$&$2.2$\\
&$E/\iota$&1.44 (input)&&$ 7 ~:~J/\psi\rar\gamma\iota$&\\
&&&&&\\
\hline 
\end{tabular}
\caption{ Unmixed gluonia masses and couplings from QSSR.}
\end{center}
\end{table}
\nin
Our results
satisfy the mass hierarchy $M_S<M_P\approx M_T$, which suggests that the
scalar meson is the lightest gluonium state. However, 
the consistency of the different sum rules in the scalar sector requires
the existence of a low mass and broad
$\sigma$-meson coupled strongly both to gluons and
to pairs of Goldstone bosons, whose effects can be
missed in a one-resonance parametrization of the spectral function, and
in the present lattice calculations. One should also notice that the values of $\sqrt{t_c}$,
which are approximately the mass of the next radial excitations, indicate that the
mass-splitting between the ground state and the radial excitations is relatively much
smaller ($30\%$) than in the case of ordinary hadrons (about $70\%$ for the $\rho$ meson), such
that one can expect rich gluonia spectra in the vicinity of 2--2.2 GeV, in addition to the
ones of the lowest ground states.\\
$\b$ We have also computed the masses 
and decay constants of the scalar quarkonia
(section 6).\\
$\b$ We have used some low-energy theorems (LET) and/or three-point function 
sum rules
in order to predict some decay widths of the bare unmixed states 
(sections 5 and 6), where the results are summarized in Table 2
\footnote{We have used the fact that the $G$ couplings to
$\pi\pi$ and $KK$ are negligible as indicated by the GAMS data \cite{GAMS}}.
\begin{table}[H]
\begin{center}
\begin{tabular}[h]{ c c c c c c c c}
\hline 
 & & & & & & &\\ 
Name&Mass [GeV]&$\pi^+\pi^-$ [GeV]&$K^+K^-$ [MeV]&$\eta\eta$
[MeV]&$\eta\eta'$ [MeV]&$(4\pi)_S$ [MeV]&$\gamma\gamma$ [keV]\\ &&&&&&\\
\hline 
&&&&&&&\\
$\sigma_B$&$0.75\sim 1.0$&$ 0.2\sim 0.5$&$SU(3)$&$SU(3)$&&&$0.2\sim 0.3$\\
&(input)&&&&&\\
$\sigma'_B$&1.37&$0.5\sim 1.3$&$SU(3)$&$SU(3)$&&$43\sim 316$&$0.7\sim 1.0$\\
&(input)&&&&&(exp)&\\
$G$&1.5&$\approx 0$&$\approx 0$&$1.1\sim 2.2$&$5\sim 10$&$60\sim 138$&$1.0\pm 0.8$\\
&&&&&&&\\
&&&&&&&\\
$S_2$&1.&0.12&$SU(3)$&$SU(3)$&&&0.67\\
$S'_2$&$1.3\approx \pi'$&$0.30\pm 0.15$&$SU(3)$&$SU(3)$&&&$4\pm 2$\\
$S_3$&$1.474\pm 0.044$&&$73\pm 27$&$15\pm 6$&&&$0.4\pm 0.04$\\
$S'_3$&$\approx 1.7$&&$112\pm 50$&$SU(3)$&&&$1.1\pm 0.5$\\
&&&&&&&\\
\hline
\end{tabular}
\caption{ Unmixed scalar gluonia and quarkonia decays}
\end{center}
\end{table}

\nin
$\b$ We have discussed some maximal
quarkonium-gluonium mixing schemes, in an attempt to explain the
complex structure and decays of the observed scalar mesons:\\
{\bf Below 1 GeV:} \\
We find that, a maximal mixing 
between two near-by quarkonium $S_2(\bar uu+\bar dd)$ and
gluonium $\sigma_B$ states around 1 GeV, can explain the large width of the
$\sigma(0.75\sim 1.)$, the narrownness of the $f_0(0.98)$ and 
its strong coupling to $\bar KK$ (this latter property
enables its production from the $\phi$ radiative decay). This scheme being
a QCD-based alternative
to the four-quark and $\bar KK$ molecule scenarios.\\
{\bf Above 1 GeV:} \\
{\bf --} The $f_0(1.37)$ is a superposition of two states, 
the $S'_2$ radial excitation of the $S_2\equiv (\bar uu+\bar dd)$
quarkonium state and the $f_0(1.37)$ coming from a maximal 
mixing between the radial excitation 
$\sigma'_B$ of broad low-mass $\sigma$ with the quarkonium
$S_3$ and gluonium $G$.\\
{\bf --} The $f_0(1.5)$ satisfying the properties observed by
the Crystal Barrel collaboration \cite{BARREL}
(namely large widths into $4\pi$, $2\pi$ and $\eta\eta'$), comes also
from a maximal 
mixing between the radial excitation 
$\sigma'_B$ of a broad low-mass $\sigma$ with the quarkonium
$S_3$ and gluonium $G$ (orthogonal partner of the $f_0(1.37)$). Our
approach also suggests that the $G(1.6)$ seen earlier by the GAMS
collaboration is likely an ``almost" pure gluonium state as emphasized
earlier in NV.\\
{\bf --} The $f_J(1.7)$ (if its spin is confirmed to be zero) can
be identified with the $S'_3$ radial excitation of the $S_3(\bar ss)$ state. 
The dip found in the $\bar KK$ mass distribution by the Crystal Barrel
and $\bar pp$ annihilation at rest \cite{ABELE,GASTALDI} 
around $(1.5\sim 1.6)$ GeV can result
from a  destructive mixing between the $f_J(1.71)$ and its
orthogonal partner, which is very wide.\\
$\b$ In the tensor sector, using a QSSR evaluation
of the off-diagonal quark-gluon two-point correlator, one finds,
 that the quarkonium-gluonium-mass mixing angle is
small, of the order of $10^\circ$ \cite{BRAMON}, which can exclude
the identification of the $f_J(1.71)$ as a $2^{++}$ gluonium
\footnote{A further improvement of our mass prediction is under way.}, but
favours the gluonium nature of the
observed $\zeta(2.2)$, where the total width satisfies our 
upper bound (section 8). However, due to the small value of the QCD continuum
threshold, which is about the mass squared of the radial excitation,
we expect to have a rich population of $2^{++}$ gluonia in this 
2 GeV region. Our result should stimulate a further search of these states and
a test of the (non-)existence of the $g_T$ seen earlier by the BNL collaboration
in the OZI suppressed $\pi^- p\rar \phi\phi n$ reaction, but not observed in
previous $J/\psi$ radiative decay data \cite{PDG,HITLIN}.
\\
$\b$ In the pseudoscalar sector, the quarkonium-gluonium mass-mixing angle 
is also small (about 12$^\circ$ \cite{PAK}),
which combined with the decay widths of the $\eta'(0.96)$ allows us to predict
the gluonium into $\gamma\gamma$ width and radiative decays (section 9). 
Finally, we found that the $E/\iota(1.44)$
is weakly coupled to the gluonic current, which can favour its
interpretation as the radial excitation of the $\eta'(0.96)$.
\vfill\eject
\section*{Acknowledgements}
Some parts of this paper are an update and improvements of the previous works based 
on QCD spectral sum rules
and low-energy theorems done by the author and his following collaborators:
Emili Bagan, Albert Bramon, Gerard Mennessier, Sonia Paban,
Namik Pak, Nello Paver, Jos\'{e} Latorre
 and Gabriele Veneziano in the years 1982 to 89. Its write-up
has been provoked by the renewed interests on the gluonia
phenomenology during the Gluonium 95 Workshop
(Propriano, Corsica), the QCD96 Euroconference
(Montpellier) and the ICHEP 96 Conference (Varsaw), which need a much more complete
analysis of the gluonia sectors than presently available works.  
It is a pleasure to thank Gabriele Veneziano for his assistance and 
for numerous discussions during the preparation  of this work. Informative 
discussions with Ugo Gastaldi
and Rolf Landua on the experimental data are also acknowledged. 
This work has been completed
during my visit at the CERN Theory Division, which I also thank for
the hospitality.


\end{document}